\documentclass[5p,twocolumn,10pt,sort&compress]{elsarticle}

\usepackage{amsmath,bm,amssymb,latexsym,galois,amsthm} 
\usepackage{caption,subcaption}
    \theoremstyle{plain}
    \theoremstyle{definition}
    \newtheorem{definition}{Definition}
    \newtheorem{theorem}{Theorem}
    \newtheorem{lemma}{Lemma}
    \newtheorem{remark}{Remark}
    \newtheorem{example}{Example}
    \newtheorem*{defn*}{Definition}
    \newtheorem{corollary}{Corollary}
    \def\eproof{{\mbox{}\hfill\qed}\medskip}
    \theoremstyle{remark}

\begin{document}

\begin{frontmatter}



\title{Quantum memory error correction computation based on Chamon model}


\author[inst1,inst2,inst3]{Jian Zhao}
\author[inst1,inst2,inst3,inst4]{Yu-Chun Wu\corref{cor1}%
}
\ead{wuyuchun@ustc.edu.cn}
\author[inst1,inst2,inst3,inst4,inst5]{Guo-Ping Guo}

\cortext[cor1]{Corresponding author}

\affiliation[inst1]{Key Laboratory of Quantum Information, Chinese Academy of Sciences, School of Physics, University of Science and Technology of China, Hefei, Anhui, 230026, P. R. China}
\affiliation[inst2]{CAS Center For Excellence in Quantum Information and Quantum Physics, University of Science and Technology of China, Hefei, Anhui, 230026, P. R. China}
\affiliation[inst3]{Hefei National Laboratory, University of Science and Technology of China, Hefei 230088, P. R. China}

\affiliation[inst4]{Institute of Artificial Intelligence, Hefei Comprehensive National Science Center, Hefei, Anhui, 230088, P. R. China}
\affiliation[inst5]{Origin Quantum Computing Hefei, Anhui 230026, P. R. China}

\begin{abstract}
Quantum error correction codes play a central role in the realisation of fault-tolerant quantum computing. Chamon model is a 3D generalization of the toric code. The error correction computation on this model has not been explored so far. In this work, the Chamon model is turned to a non-CSS error correction code. Logical qubits are built by the construct of logical Pauli operators. The property of logical operators reveals the expressions of code distance. According to the topological properties of Chamon models, an error elimination algorithm is proposed. Based on the error elimination algorithm, we propose a global randomized error correction algorithm to decode Chamon models in every single-qubit depolarized channel. This decoding algorithm is improved by adding the pretreatment process, termed the probabilistic greedy local algorithm, which adapts to different kinds of high-dimensional models. The estimated threshold error rate for numerical experiment can be raised to $4.92\%$.
\end{abstract}


%

\end{frontmatter}


\section{Introduction}
When processing certain computing tasks, it is reasonable to presume that the potential effect of quantum computers may outperform classical computers\cite{biamonte2017quantum}.
Quantum computing promises exponentially or quadratically faster
processing of factoring problems\cite{shor1994algorithms,monz2016realization}, quantum simulating problems \cite{lloyd1996universal} and searching problems\cite{grover1996fast}.
However, due to the objective existence of quantum noise, the realization of practical quantum computers suffers challenge.

Quantum error correction codes play a central role in the realisation of fault-tolerant quantum computing\cite{terhal2015quantum,roffe2019quantum}. The type of encoding determines the arrangement of qubits in the hardware and the control of quantum gates at the software level.
Many errors correcting codes have been realised, including Shor code\cite{luo2021quantum}, surface code\cite{andersen2020repeated,ai2021exponential,erhard2021entangling}, color code\cite{nigg2014quantum,ryan2021realization}, repetition code\cite{chiaverini2004realization,schindler2011experimental,wootton2018repetition} and bosons quantum error-correcting code\cite{campagne2020quantum}.
The work of high-dimensional error-correcting codes mainly stays at the theoretical level, such as $4$-D toric code\cite{dennis2002topological}, high-dimensional color code\cite{castelnovo2008topological} and cubic code\cite{bravyi2011analytic}.
These proposed high-dimensional error correction codes with geometric topology structures belong to CSS codes, whose stabilizers can be divided into two classes.

A natural question is whether it is possible to analyse fault-tolerant quantum computation in non-CSS codes with topological structures.
The $XZZX$ code as a variant of the surface code has been proposed \cite{bonilla2021xzzx}.
We turn attention to high-dimensional non-CSS code out of curiosity.

The original Chamon model is proposed to present a many-body Hamiltonian of the system that cannot reach their ground states when the environment temperature is closed to absolute zero, termed topologically ordered ground states\cite{chamon2005quantum}. 
It can be regarded as a $3$-dimensional generalization of toric codes with six-qubit nearest-neighbor interactions on a $3$-dimensional lattice.
Bravyi investigated thoroughly the types of the excitations in this model and demonstrated the ground state degeneracy\cite{bravyi2011topological}. 
Recently, Shirley has studied the fermionic gauge theory and fracton topological order in Chamon models\cite{shirley2023emergent}.
Nevertheless, it has not been considered so far for error correction quantum computation based on Chamon models.
For filling in gaps and inspired by previous work, we turn Chamon model to a kind of error correction code.

In this paper, a construction method of logical qubits is proposed in Chamon model with arbitrary scales. The expression of code distance in Chamon models is derived.
According to the topological properties of Chamon models, this paper presents
a specific error correction algorithm. 
An improved error correction algorithm is also proposed, which slightly raise the threshold to $4.92\%$.

For given $N$ data qubits, one can construct a stabilizer code with $[N,k,d]$, where $k$ is the number of constructed logical qubits and $d$ is called code distance\cite{nielsen2002quantum}.
The sizes of $k$ and $d$ respectively reflect the ability to accommodate the number of logical qubits and the ability to fault tolerance\cite{knill2000theory}.
In Chamon models, there exists a trade-off between $k$ and $d$. 
Its asymptotic properties are manifested that when the number of physical qubits is given $\mathcal{O}(n^3)$, the maximum number of logical qubits can reach $\mathcal{O}(n)$, with $d=\mathcal{O}(n)$; if the number of  logical qubits are sacrificed to a constant, the code distance $d$ can reach $\mathcal{O}(n^2)$. For example, if encoding constant logical qubits, $\mathcal{O}(d^2)$ physical qubits are needed in surface code, but in Chamon models we only need to use $\mathcal {O}(d^{3/2})$ qubits.

Theoretically, the code distance in Chamon models with dimensions $\alpha_x, \alpha_y, \alpha_z$ is stil an unresolved problem. For Chamon models with $N$ qubits, the code distance $d$ was estimated by a rough upper bound $\mathcal O(\sqrt{N})$\cite{bravyi2011topological}. We solved this problem in two cases that the code distance $d=\min\{\alpha_y \alpha_z,\alpha_z \alpha_x,\alpha_x \alpha_y\}$ when $\alpha_x,\alpha_y,\alpha_z$ are pairwise coprime and $d=2\alpha$ when $\alpha_x,\alpha_y,\alpha_z=\alpha$. 

The discussion of paper is organized as follows. In Sec.\ref{sec_Chamon}, the definition of Chamon model is introduced.
In Sec.\ref{sec_logical}, by the constructions of logical operators, Chamon model is turned to $4\alpha$ logical qubits and the properties of logical operators are derived. 
The code distances of different Chamon models are analysed in Sec.\ref{sec_distance}. 
When three dimensions are coprime, the logical operators become half-chain operators forming a loop in toric representations, which is put in \ref{sec_appendix_torus}.
In Sec.\ref{sec_decode}, for decoding Chamon models, the global error correction algorithm is constructed and the threshold is estimated, $p_{th}=4.45\%$.
Consiering local properties of sparse qubits errors, the decoding algorithm is improved, such that the threshold becomes $4.92\%$.
To organize the material, the proof of Lemma \ref{lem_cijinlin} is put in \ref{sec_appendix_lemma} and the procedure of the probabilistic greedy local algorithm is expounded in \ref{sec_appendix_greedy}.
Last in Sec.\ref{sec_conclusion}, we draw the conclusions of this paper.

\textbf{Notation.}
The paper uses capital Roman letters $A$, $B$,$\ldots$, for matrices or operators, lower case Roman letters $x$, $y$,$\ldots$, for vectors,
and Greek letters $\alpha$, $\beta$,$\ldots$, for scalars. In three-dimensional space, denote the unit vectors by $e_x^{\pm}=(\pm1,0,0)$, $e_y^{\pm}=(0,\pm1,0)$, $e_z^{\pm}=(0,0,\pm1)$. 
Specially, denote the Pauli operators by $X,Z$, and $Y=ZX$.

\section{Chamon model}\label{sec_Chamon}
In this section, first introduce the definition of Chamon model, then analyse the 
degrees of freedom in code space. 
The degrees of freedom  determines the maximum number of logical qubits used.
\subsection{Building Chamon model}
On three-dimensional cubic Bravais lattice, qubits are placed on the face centers each neighbouring two cubes and the vertices each shared in eight cubes. On each site there is only one qubit.
The stabilizer acts on six qubits, which is closest to either the midpoint of the fixed side of a cube or the body center of a cube. According to the directions of the cube side, called $x$-direction,$y$-direction and $z$-direction, these six qubits are divided into three classes. Each class includes two qubits, whose sites connecting a line parallel to one of the three directions.
The stabilizers action type is corresponding $X,Y$ and $Z$.

For convenience and rigor, it is necessary to introduce some notations. Given a positive integer $\alpha$, the set $\mathcal Z_\alpha=\{0,1,\dots,\alpha-1\}$ equipped with a modulo-$\alpha$ addition operation forms a group. The Chamon model is constructed by the group $\mathcal A=\mathcal Z_{2\alpha_x}\times\mathcal Z_{2\alpha_y}\times\mathcal Z_{2\alpha_z}$.

\begin{definition}
    Given a group $\mathcal A=\mathcal Z_{2\alpha_x}\times\mathcal Z_{2\alpha_y}\times\mathcal Z_{2\alpha_z}$, $\alpha_x,\alpha_y,\alpha_z\in\mathcal Z^{+}$, let $\mathcal D=\{(x,y,z)\in\mathcal A|(x+y+z)/2\in\mathcal Z\}$.
    Data qubits are placed on each point of $\mathcal D$, called data qubits set. For all $s=(s_x,s_y,s_z)\in\mathcal A\backslash\mathcal D$, denote $s+e^{\pm}$ briefly by $e^{\pm}$, then the stabilizers group is generated by
    \begin{align}\label{eq_S_d}
    S_s=X_{e_x^{+}}X_{e_x^{-}}Y_{e_y^{+}}Y_{e_y^{-}}Z_{e_z^{+}}Z_{e_z^{-}}.
    \end{align}
    Denote the stabilizers group by $\mathcal S_{\mathcal D}=\langle S_s\rangle$, $s\in\mathcal A\backslash\mathcal D$. 
    Then the data qubits set $\mathcal D$ and the corresponding stabilizers group $\mathcal S_{\mathcal D}$ form the Chamon model.
\end{definition}

\begin{figure}[h]
  \centering
   \includegraphics[width=5cm]{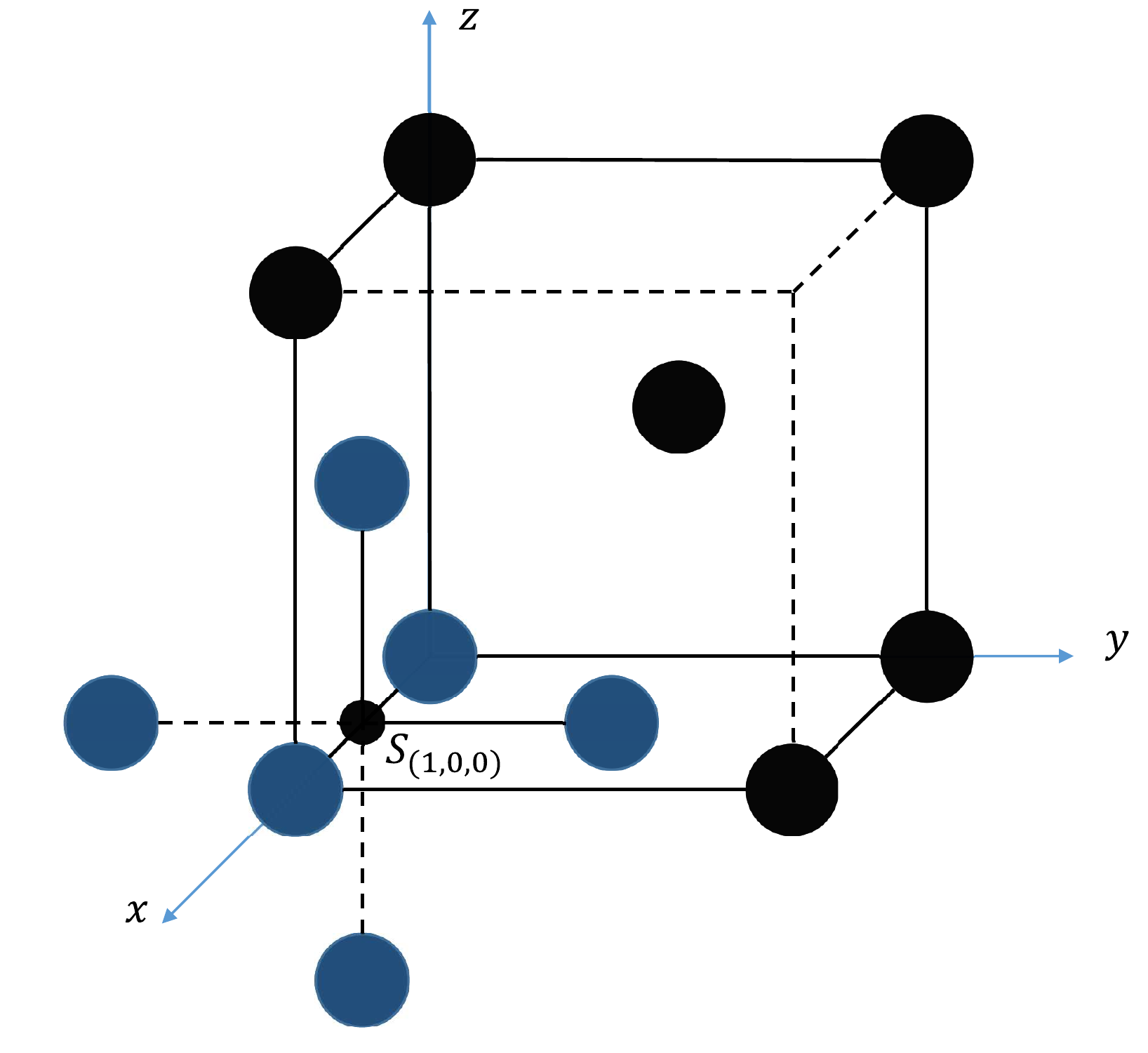}\\
  \caption{{\bf The Chamon model.}  The stabilizer $S_{(1,0,0)}$ acts six blue qubits. Data qubits are placed on the face centers and the vertices  of the cubes. Each side of the cubes represents one stabilizer.  To be seen clearly some data qubits are omitted. }\label{Fig_definition}
\end{figure}

In Chamon model, each qubit maps to a stabilizer one by one according to the definition. Specifically, measure qubits can be put a unit length from every data qubit along the positive x-axis; see Fig.\ref{Fig_definition}. 
\begin{remark}
In the special case when $\alpha_y=1$, all the stabilizers in Chamon model $S_s$ are coupled with only four data qubits. The reason is the two coupled qubits in $Y$-direction are in fact the same for each fixed stabilizer. Hence $Y_{e_y^{+}}=Y_{e_y^{-}}$. Then the stabilizers are degenerated to an interesting case:
\begin{align*}
S_s=X_{e_x^{+}}X_{e_x^{-}}Z_{e_z^{+}}Z_{e_z^{-}},
\end{align*}
which is exactly regarded as the two-layer surface code independently with $XZZX$ type stabilizers \cite{bonilla2021xzzx}. So is the case that $\alpha_x=1$ or $\alpha_z=1$ due to the rotation symmetry. For the rest of this paper, only consider general situations where $\alpha_x,\alpha_y,\alpha_z>1$.
\end{remark}

The definition of stabilizers is rational. 
Note that all the stabilizers always commute. 
It is trivial that $S_s$ and $S_{s^{\prime}}$ do not have qubits shared. Once the stabilizers share qubits in common, they  either share only one data qubit applied to the same acting type, in which case they commute apparently, or share two qubits, in which case the operators acting on each shared qubit anti-commute. Thus for all the stabilizers, $S_sS_{s^{\prime}}$=$S_{s^{\prime}}S_s$.

\subsection{Analysing the Eigenspace}
Chamon model as a kind of stabilizer code is a subspace of code space $\mathcal H^{\otimes |\mathcal D|}$. 
For any quantum state $|\psi\rangle\in \mathcal H^{\otimes |\mathcal D|}$, all stabilizers $S_s$ force the Hilbert space into a common eigen-subspace. Based on the measurement results obtained by the stabilizers, the initial state is collapsed on a fixed eigenstate. 
For lack of consideration for the independence of the generated stabilizers, the number of stabilizers in Eq.\eqref{eq_S_d} is $|\mathcal D|=4\alpha_x\alpha_y\alpha_z$, $i=0,\dots,|\mathcal D|-1$. 
The measurement outcomes can be denoted by a $|\mathcal D|$ dimension two-valued vector $r=(r_0,\dots,r_{|\mathcal D|-1})$, $S_i|\phi\rangle=r_i|\phi\rangle$, $r_i\in\{-1,1\}$. Note that the label $S_i$ has been remarked.  
For a fixed measurement $r$, the eigenspace is denoted by

\begin{footnotesize}
    \begin{align*}
\mathcal E_r=\left\{|\phi\rangle\in\mathcal H^{\otimes |D|}\big|S_i|\phi\rangle=r_i|\phi\rangle, S_i\in \mathcal S_{\mathcal D},i=0,\dots,|\mathcal D|-1\right\}.
    \end{align*}
\end{footnotesize}

Denote the greatest common divisor of $\alpha_x,\alpha_y,\alpha_z$ by $\alpha$. Bravyi has proved that the dimension of the eigenspace $\mathcal E_r$ is $2^{4\alpha}$
\cite{bravyi2011topological}. Our purpose is to build $4\alpha$ logical qubits in any eigenspace.

\section{Logical qubits}\label{sec_logical}
In this section, by the constructions of proper logical operators, the logical qubits 
are built in any Chamon model.
\subsection{The special case in $\alpha=1$}\label{sec_alpha1}
When $\alpha=1$, at most Chamon model can form $4\alpha=4$ logical qubits. The purpose is to construct a kind of logical $X$ operators and logical $Z$ operators, denoted by $X_L^{(i)}$ and $Z_L^{(i)}$, satisfying
\begin{align}\label{eq_logical_qubits}
{X_L^{(i)}}^2=&{Z_L^{(i)}}^2=I\notag\\
X_L^{(i)}Z_L^{(j)}=&(-1)^{\delta_{ij}}Z_L^{(j)}X_L^{(i)},
\end{align}
where $\delta_{ij}=0$ when $i\neq j$ or $\delta_{ij}=1$ when $i=j$, $i,j=0,1,2,3$. Given an eigenspace $\mathcal E_r$, all the logical operators must commute with $S_i$, which ensures that the logical operation is performed inside the eigenspace.  Identical to Pauli relationships in data qubits, the logical $Y$ operators are defined as $Y_L^{(i)}\triangleq Z_L^{(i)}X_L^{(i)}$.

It is not unique to construct proper logical operators $X_L$ and $Z_L$. A naive thought is the constructed $X_L^{(i)}$ can be generated by only bit flips $X$ on data qubits, $Z_L^{(i)}$ generated by only phase flips $Z$.
Next, introduce the construction process of the logical operators through classifications of data qubits.
Note that the qubits in data qubits set $\mathcal D$ can be divided into four subsets, which contain  one vertex qubits set and 3 face-centered qubits sets. These qubits sets can be generated by the points $g_0=(0,0,0)$, $g_1=(1,1,0)$, $g_2=(1,0,1)$ and $g_3=(0,1,1)$ respectively.
Define the data qubits set $\mathcal D_x^{g_i},\mathcal D_z^{g_i}\subset \mathcal D$ as
\begin{align}\label{eq_D_half}
\notag\mathcal D_x^{g_i}=\{d|d=g_i+2\lambda_x e_z^{+}+2\lambda_y e_y^{+},\lambda_x,\lambda_y\in\mathcal Z\}\\
\mathcal D_z^{g_i}=\{d|d=g_i+2\lambda_x e_x^{+}+2\lambda_y e_y^{+},\lambda_z,\lambda_y\in\mathcal Z\},
\end{align}
where $i=0,1,2,3$. Clearly, the intersection of $\mathcal D_x^{g_i}$ and $\mathcal D_x^{g_j}$ is empty for $i\neq j$. The union of $\mathcal D_x^{g_0}$ and $\mathcal D_x^{g_3}$ is $\{d\in \mathcal D|d_x=0\}$, denoted by $\mathcal D_{x=0}$ and $\mathcal D_x^{g_1}\cup D_x^{g_2}=\mathcal D_{x=1}$, both constituting a lattice plane parallel to $yOz$ plane, respectively. The same considerations apply to $\mathcal D_z^{g_i}$.
Define $X|_{\mathcal D^{\prime}}=\prod _{j\in\mathcal D^{\prime}}X_j$, then 

\begin{theorem}\label{Thm_g=1}
Given a Chamon model with $\alpha=1$, the operators $X|_{\mathcal D_x^{g_i}}$ and $Z|_{\mathcal D_z^{g_i}}$ make the Chamon model $4$ logical qubits.
\end{theorem}
\proof
Firstly, for a fixed $i$, $X^2|_{\mathcal D_x^{g_i}}=I=Z^2|_{\mathcal D_z^{g_i}}$. Note that $\mathcal D_x^{g_i}\cap\mathcal D_z^{g_i}=\{d|d=g_i+2\lambda_y e_y^{+},\lambda_y\in\mathcal Z\}$, which contains $\alpha_y$ elements. Based on the assumption that $\alpha_y$ is odd, hence the operators $X|_{\mathcal D_x^{g_i}}$ and $Z|_{\mathcal D_z^{g_i}}$ anti-commute. Since all the stabilizers commute with each other, $X|_{\mathcal D_x^{g_i}}$ and $Z|_{\mathcal D_z^{g_i}}$ cannot be generated by stabilizers. One can easily check $X|_{\mathcal D_x^{g_i}}$ and $Z|_{\mathcal D_z^{g_i}}$ commute with each stabilizer generator. Thus the Chamon model can be used as logical qubits.
Further since $D_x^{g_i}\cap D_z^{g_j}=\varnothing$, the operators $X|_{\mathcal D_x^{g_i}}$ and $Z|_{\mathcal D_z^{g_j}}$ pairwise commute, $\forall i\neq j$. It implies the operators satisfy the conditions in Eq.\eqref{eq_logical_qubits}. The commutation relations hold the independence of logical operators $X|_{\mathcal D_x^{g_i}},Z|_{\mathcal D_z^{g_i}}$, that is they cannot be generated by other logical operators $X|_{\mathcal D_x^{g_j}},Z|_{\mathcal D_z^{g_j}}$, $i\neq j$. We have shown the Chamon model indeed forms 4 logical qubits.

\eproof


For analysing the property of logical operators here introduce some concepts. Consider the centralizer of $\mathcal S$ in Pauli group denoted by
\begin{align*}
\mathcal C(\mathcal S)=\langle \mathcal S,X|_{\mathcal D_x^{g_i}},Z|_{\mathcal D_z^{g_i}}\rangle,
\end{align*}
where $i=0,1,2,3$. Let us concern more about all the logical operators with $\mathcal S$ invariance. Thus introduce the quotient group generated by
\begin{align*}
\mathcal C(\mathcal S)/\mathcal S=\langle  X|_{\mathcal D_x^{g_i}}, Z|_{\mathcal D_z^{g_i}}\rangle,
\end{align*}
where denote the equivalence class of logical operators by the same notations.

Note that the logical operators can be translated. For example, for $g_0=(0,0,0)$ and $g_0^\prime=(2,0,0)$, clearly,  $X|_{\mathcal D_x^{g_0}}=X|_{\mathcal D_x^{g_0^\prime}}$, due to 
\begin{align*}
  \prod_{S_j\in\mathcal S^\prime}S_j X|_{\mathcal D_x^{g_0}}=X|_{\mathcal D_x^{g_0^\prime}},
  \end{align*}
where $\mathcal S^\prime=\{S_s\in \mathcal S| s=e_x^{+}+2\lambda_y e_y^{+}+2\lambda_z e_z^{+},\lambda_y,\lambda_z\in\mathcal Z \}$. Thus, It can be concluded that the logical operators $X|_{\mathcal D_x^{g_i}}, Z|_{\mathcal D_z^{g_i}}$ can be translated across the x-axis and z-axis directions, respectively.

Coupled with half of the qubits on a certain plane, the operators $X|_{\mathcal D_x^{g_i}}$ and $Z|_{\mathcal D_z^{g_i}}$ are termed half-plane operators. It is clear that translated half-plane operators defined in Thm.\ref{Thm_g=1} are still half-planar. 

\subsection{The general case}
In this section, the logical $X$ operators and logical $Z$ operators in $\mathcal C(\mathcal S)/\mathcal S$ are introduced  generally.

\begin{figure}[ht]
  \centering
   \includegraphics[width=0.8\linewidth]{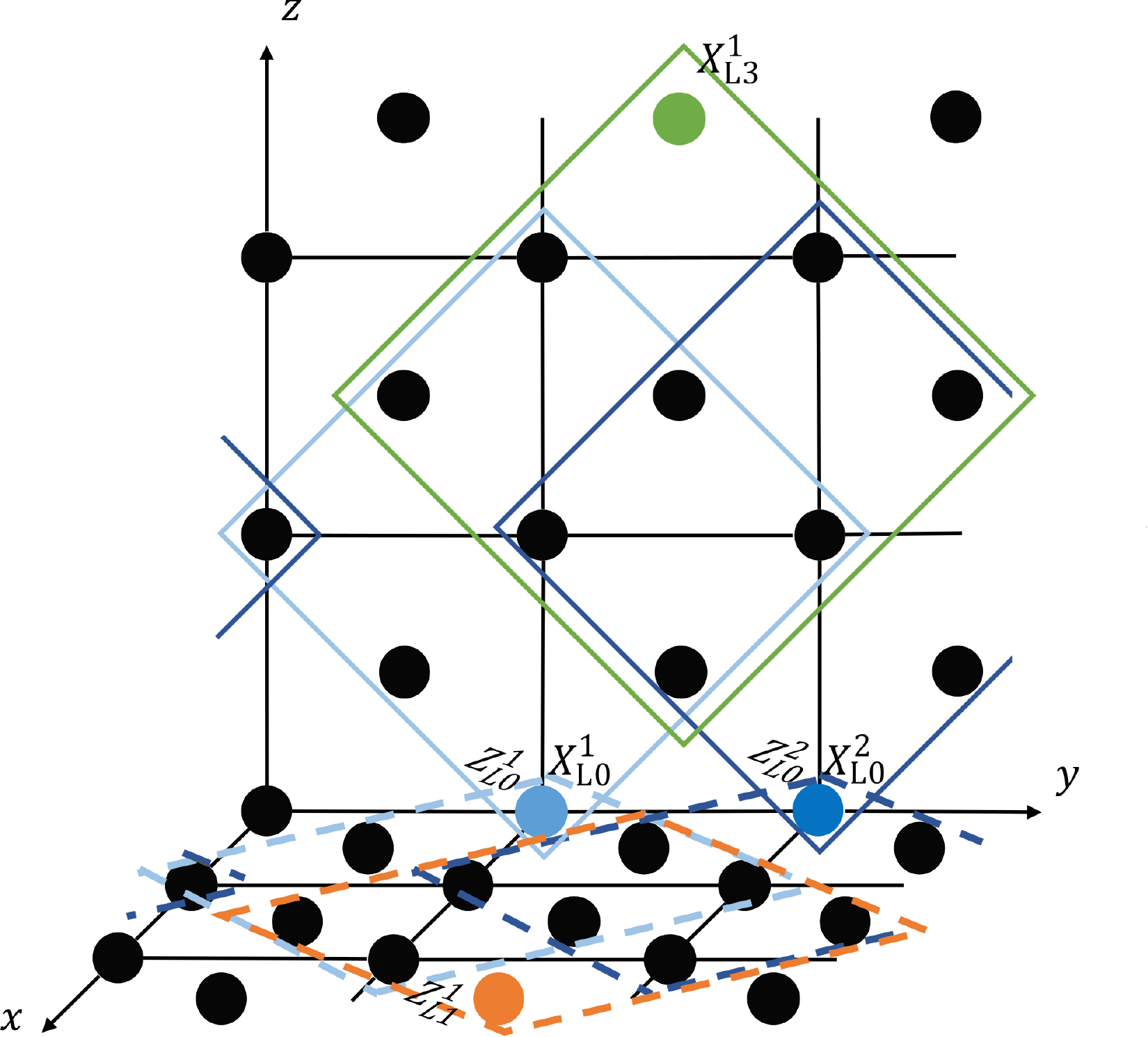}\\
   \vspace{0.3em}
   \includegraphics[width=0.8\linewidth]{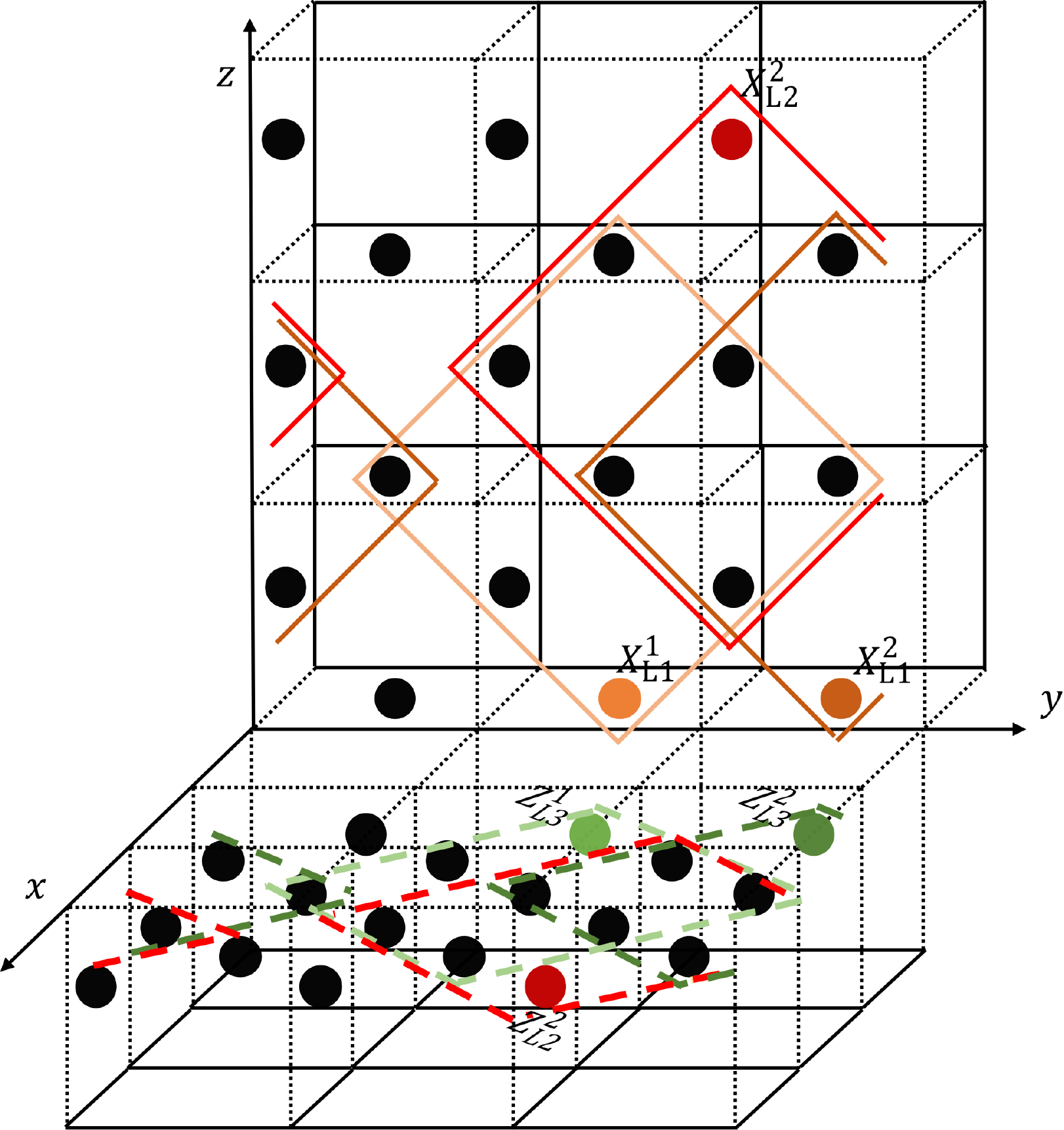}
  \caption{{\bf The Chamon model with $\alpha_x,\alpha_y,\alpha_z=3$.}  Data qubits are set with square borders in different colors, corresponding to the logical operators. Among them, the blue border corresponds to the case $i=0$, a lighter blue for $j=1$, darker for $j=2$. Also, the sites $g_{i,j}$ are filled with the corresponding colors. Solid borders denote the logical $X$ operators and dotted borders denote the logical $Z$ operators. The part of the logical operators is still omitted. }\label{Fig_XLZL}
\end{figure}

Before the constructions, we first visualize the geometry configurations of the logical operators by an example with $\alpha_x=\alpha_y=\alpha_z=3$. The greatest common divisor is $\alpha=3$. Let $g_{0,1}=(0,2,0)$ and define a square half-plane operator in the $yOz$ plane as
\begin{small}\begin{align*}
\mathcal D_x^{{0,1}}=\{d|d=g_{0,1}+\lambda(e_z^{+}+e_y^{+})+\lambda^{\prime}(e_z^{+}+e_y^{-}),\lambda,\lambda^{\prime}\in\mathcal Z_\alpha\}.
\end{align*}\end{small}
Then apply $X$ operator on each qubit located in $\mathcal D_x^{{0,1}}$, denoted by $X|_{\mathcal D_x^{{0,1}}}$. It is coupled with $9$ qubits, which form a square occupying half of the $yOz$ plane. One can check it commutes with all stabilizers; see the operator $X_{L0}^1$ in Fig.\ref{Fig_XLZL}.

Applying this idea to the case when $\alpha_x=\alpha_y=\alpha_z=\alpha$,  define $g_{i,j}=g_i+2je_y^{+}$, and let
\begin{footnotesize}\begin{align}\label{eq_alphax=y=z}
\notag\mathcal D_x^{i,j}=&\{d|d=g_{i,j}+\lambda(\mu_{ij}e_z^{+}+e_y^{+})+\lambda^{\prime}(\mu_{ij}e_z^{+}+e_y^{-}),\lambda,\lambda^{\prime}\in\mathcal Z_\alpha\},\\
\mathcal D_z^{i,j}=&\{d|d=g_{i,j}+\lambda(\nu_{ij}e_x^{+}+e_y^{+})+\lambda^{\prime}(\nu_{ij}e_x^{+}+e_y^{-}),\lambda,\lambda^{\prime}\in\mathcal Z_\alpha\},
\end{align}\end{footnotesize}where $i=0,1,2,3$ and $j=0,1,2,\dots,\alpha-1$, $\mu_{ij}=(-1)^{[i/2]}, \nu_{ij}=(-1)^{\lceil i/2\rceil({\rm mod }\hspace{0.2em}2)}$. Let $X_{Li}^j=X|_{\mathcal D_x^{i,j}}$ and $Z_{Li}^j=Z|_{\mathcal D_x^{i,j}}$,  termed
square logical operators.

\begin{lemma}\label{Lem_alphax=alphayz}
Given a Chamon model with $\alpha_x=\alpha_y=\alpha_z=\alpha$, the operators $X_{Li}^j$ and $Z_{Li}^j$ make the Chamon model $4\alpha$ logical qubits.
\end{lemma}
\proof
Due to $D_x^{i,j}\cap D_z^{i,j}=\{g_{i,j}\}$, it implies that the operators $X_{Li}^j$ and $Z_{Li}^j$ anti-commute. And these square half-plane operators commute with all the stabilizers. 
Lastly to verify that $X_{Li}^j$ and $Z_{Li^{\prime}}^{j^{\prime}}$ commute,  one can  check $D_x^{i,j}\cap D_z^{i^{\prime},j^{\prime}}=\varnothing$, $\forall (i,j)\neq ( i^{\prime},j^{\prime})$. 
Thus, $X_{Li}^j$ and $Z_{Li}^j$ as logical operators make the Chamon model $4\alpha$ logical qubits.
\eproof

A specific example is presented with $\alpha_x,\alpha_y,\alpha_z=3$; see Fig.\ref{Fig_XLZL}. Note that each logical operator occupies a square half-plane.

The construction of logical operators generally is available in previous ideas of square half-plane operators. Denote the greatest common divisor of $\alpha_x,\alpha_y,\alpha_z$ by $\alpha=g(\alpha_x,\alpha_y,\alpha_z)$. Then $\alpha_x=\beta_x \alpha$, $\alpha_y=\beta_y \alpha$, $\alpha_z=\beta_z \alpha$, where $\beta_x,\beta_y$ and $\beta_z$ satisfy $g(\beta_x,\beta_y,\beta_z)=1$.
There must exist an odd number in $\beta_x,\beta_y$ and $\beta_z$. Without loss of generality, suppose $\beta_y$ is odd. 
An intuitive thought is that the plane parallel to $yOz$ plane in Chamon model can be divided into $\beta_y\beta_z$  congruent squares. Apply the corresponding $X$ operators on every square and find they constitute logical $X$ operators; similarly, it is applied to logical $Z$ operators.

Based on the previous analysis, it is necessary for applying to the general case to modify the definition in Eq.\eqref{eq_alphax=y=z}. Let
\begin{small}\begin{align}
D_x^{i,j}=\{d|d=g+\lambda(\mu_{ij}e_z^{+}+e_y^{+})+\lambda^{\prime}(\mu_{ij}e_z^{+}+e_y^{-}),\notag\\\lambda,\lambda^{\prime}\in\mathcal Z_\alpha, g\in \mathcal G_{i,j}^x\},\\
\mathcal D_z^{i,j}=\{d|d=g+\lambda(\nu_{ij}e_x^{+}+e_y^{+})+\lambda^{\prime}(\nu_{ij}e_x^{+}+e_y^{-}),\notag\\\lambda,\lambda^{\prime}\in\mathcal Z_\alpha, g\in \mathcal G_{i,j}^z\},
\end{align}\end{small}
where
\begin{align*}
\mathcal G_{i,j}^x=&\{g|g=g_{i,j}+2\lambda\alpha e_y^{+}+2\lambda^{\prime}\alpha e_z^{+},\lambda\in\mathcal Z_{\beta_y},\lambda^{\prime}\in\mathcal Z_{\beta_z}\},\\
\mathcal G_{i,j}^z=&\{g|g=g_{i,j}+2\lambda\alpha e_y^{+}+2\lambda^{\prime}\alpha e_x^{+},\lambda\in\mathcal Z_{\beta_y},\lambda^{\prime}\in\mathcal Z_{\beta_x}\}.
\end{align*}

\begin{theorem}\label{general}
Given a Chamon model with $\alpha=g(\alpha_x,\alpha_y,\alpha_z)$, the operators $X_{Li}^j$ and $Z_{Li}^j$ make the Chamon model $4\alpha$ logical qubits.
\end{theorem}


Let us go on to study other properties of square logical $X,Z$ operators.
Firstly, observe the relation of the nearest neighbor square operators. In Fig.\ref{Fig_XLZL}, $X_{L0}^1$ and $X_{L3}^1$ share 6 qubits and the support of $X_{L3}^1$ can be translated from $X_{L0}^1$, that is
\begin{align}\label{eq_31}
\mathcal D_x^{3,1}=\mathcal D_x^{0,1}+e_z^{+}+e_y^{+},
\end{align}
where $\mathcal D'=\mathcal D+x$ indicates $\mathcal D'=\{d+x, d\in\mathcal D\}$.
It implies $X_{L0}^1 X_{L3}^1=X|_{\mathcal D}$, $\mathcal D=\{d|d=g_{0,1}+\lambda(e_z^{+}+e_y^{-}),\lambda\in\mathcal Z_{2\alpha}\}$, which forms a rigid chain of length $2\alpha$. Generally,

\begin{corollary}{\label{Co_rigid}}
There exists a logical $X$ operator forming a rigid chain of length $2\beta_y\beta_z\alpha$, similarly, the logical $Z$ operator forming a rigid chain of length $2\beta_y\beta_x\alpha$.
\end{corollary}

Firstly for any square $X$ operator in the plane parallel to $yOz$, explain it is a logical operator and give the expression generated by $X_{Li}^j$. 
Let
\begin{align}\label{eq_11}
\notag\mathcal D_x^{0,2}=&\mathcal D_x^{0,1}+2e_y^{+},\\
\mathcal D'\triangleq& \mathcal D_x^{0,1}+e_z^{-}+e_y^{+}.
\end{align}
One can check $X_{L0}^2 X|_{\mathcal D'}=X_{L0}^1 X_{L3}^1$ because they represent the same rigid chain. It implies that the square operator $X|_{\mathcal D'}$ can be generated by other 3 logical operators, $X|_{\mathcal D'}=X_{L0}^1 X_{L3}^1 X_{L0}^2$. Define the square operators as $X_{L0}^1,X_{L3}^1,X_{L0}^2 X|_{\mathcal D'}$ the adjacent $4$ square operators, the support of which satisfies the adjacent relations in Eq.\eqref{eq_31}\eqref{eq_11}. Thus, in the $yOz$ plane, for any square $X$ operator, it can always be generated by $X_{L0}^j,X_{L3}^j$, $j=0,\dots,\alpha-1$. In other words, the product of any adjacent $4$ square operators is unit operator with $\mathcal S$ invariance.
Take the square half-plane logical X operator as an example, and define  square half-plane operator as $X_{sq}(g)=X|_{\mathcal D_x^{\mu_l\nu_l}(g)}$, 
\begin{small}\begin{align*}
  \mathcal D_x^{\mu_l\nu_l}(g)=\{d|d=g+\lambda\mu_l( e_z^{+}+e_y^{+})+\lambda^{\prime}\nu_l( e_z^{+}+e_y^{-}),\\\lambda,\lambda^{\prime}\in\mathcal Z_\alpha, g\in \mathcal D\},
\end{align*}\end{small}
where $\mu_l,\nu_l\in\{1,-1\}$.
In the fact, the whole relations of adjacent $4$ square operators are as follows,
\begin{corollary}{\label{Co_adjacent4}}
  With $\mathcal S$ invariance, 
  \begin{small}\begin{align}\label{eq_4sqrare_ver_I}
    X_{sq}(s+e_x^{+})X_{sq}(s+e_x^{-})X_{sq}(s+e_y^{+})X_{sq}(s+e_y^{-})=I\notag,\\
    X_{sq}(s+e_z^{+})X_{sq}(s+e_z^{-})X_{sq}(s+e_x^{+})X_{sq}(s+e_x^{-})=I\notag,\\
    X_{sq}(s+e_z^{+})X_{sq}(s+e_z^{-})X_{sq}(s+e_y^{+})X_{sq}(s+e_y^{-})=I,
  \end{align}\end{small}
where $s\in\mathcal A/\mathcal D$.
\end{corollary}
\proof
  The product of these operators either forms the envelope surface of the stabilizers or form identical transformation. We can check the two former transformations are exactly the product of all stabilizers in the envelope.
\eproof

This result can be extended similarly to square half-plane $Y,Z$ operators.
Further, it reveals that any square half-plane operator can be generated by $4\alpha$ square half-plane logical Pauli operators defined in Thm.\ref{general}.

Let us review the half-plane operators in the special case $\alpha=1$. Generally, when $\alpha\neq 1$, still denote the corresponding half-plane operator by $X|_{\mathcal D_x^{g_i}}, Z|_{\mathcal D_z^{g_i}}$, where $\mathcal D_x^{g_i}, \mathcal D_z^{g_i}$ have the same expressions as Eq.\eqref{eq_D_half}. The relations between half-plane operators and square logical operators are
\begin{corollary} Each half-plane operator is a logical operator and can be generated by square logical operators. Specifically,
\begin{align}\label{eq_XZhalf-square-relation}
\notag X|_{\mathcal D_x^{g_i}}=\prod_{j=0}^{\alpha-1} X_{Li}^j,\\
Z|_{\mathcal D_z^{g_i}}=\prod_{j=0}^{\alpha-1} Z_{Li}^j.
\end{align}
\end{corollary}
\proof
Consider the product of square operators $X_{L0}^j$, $j=0,\dots,\alpha-1$. The support of this product operator is constrained in the $yOz$ plane. In this plane, the data qubits in $\mathcal D_x^{g_0}$ are applied flip operations odd times. The others in $\mathcal D_x^{g_3}$ are all flipped even times, which implies $X|_{\mathcal D_x^{g_0}}=\prod X_{L0}^j$. Similarly, the situations in Eq.\eqref{eq_XZhalf-square-relation} can be proved. Note that half-plane logical operators can be translated, which means any half-plane operators can be generated by square logical operators with $\mathcal S$ invariance.
\eproof

The result shows that for any Chamon model, not only can $4$ logical qubits be constructed, but in each logical qubit space there are $\alpha$ logical qubits encoded. 

\section{The distance of Chamon model}\label{sec_distance}
In this section, introduce the code distance of Chamon models to analyze the error correction capability. And there exists a trade-off between the code distance and the number of logical qubits in one Chamon model.

\begin{definition}
For a stabilizer code $(\mathcal D, \mathcal S)$, the weight of a logical operator $L$ is defined as the minimum number of qubit bit flips or phase flips needed in $L$ with $\mathcal S$ invariance, denoted by $|L|$. Then the code distance $d$ is defined as the minimum weights of
logical operators, that is
\begin{align*}
d=\min_{L\neq I, L\in\mathcal C(\mathcal S)/\mathcal S} |L|.
\end{align*}
\end{definition}

In some cases, the explicit expressions of code distance are given.

\begin{theorem}\label{thm_d_coprime}
  Suppose $\alpha_x,\alpha_y$ and $\alpha_z$ are pairwise coprime,
  then for Chamon models with linear dimensions $\alpha_x,\alpha_y,\alpha_z$,
  the expression of code distance $d$ is described as
\begin{align}
  d=\min\{\alpha_x\alpha_y,\alpha_y\alpha_z,\alpha_z\alpha_x\}.
\end{align}
\end{theorem}

The proof of Thm.\ref{thm_d_coprime} is based on the toric representations; 
see \ref{sec_appendix_torus}.

%
%
%
%
%
%
%
%
%

\begin{theorem}
  Given a Chamon model with linear dimensions $\alpha_x=\alpha_y=\alpha_z=\alpha$, 
  the expression of code distance $d$ is described as
\begin{align}
  d=2\alpha.
\end{align}

\end{theorem}

\proof
Consider one data qubit flip, which always affects the nearest 4 stabilizers in a plane, locally. 
There are two methods to  modify the stabilizers measurements by applying flips on the nearest qubits. One is along 3 diagonal lines and the other is along coordinate axis.
Since the logical operators commute with all stabilizers, the  affected stabilizers measurements need to be eliminated,
which means the support of logical operator needs to form a loop. If the loop is along coordinate axis, the operator becomes half-plane operator with weight $\alpha^2$.

Another scenario has been discussed in Corollary \ref{Co_rigid}. One can always find the logical chain operator arranged diagonally with weight $2\alpha$.  Thus the minimum weight is $2\alpha$.
  \eproof 

  For a general Chamon model with dimensions $\alpha_x,\alpha_y$ and $\alpha_z$.
  According to the greatest common divisor $\alpha$, the logical operator generators can always be divided into fragments distributed in different squares.
  It becomes too complicated to guarantee that the logical operator with minimum weight can be regularly divided.
  However, one can easily construct a logical chain operators according to Corollary \ref{Co_rigid}, showing the code distance $d\leq \min\{2\beta_y\beta_z\alpha,2\beta_y\beta_x\alpha\}$. 

  It reveals the relationship between the number of physical qubits, the number of constructed logical qubits and the code distance, and there is a balance between these parameters. When the number of physical qubits is given, the larger the number of logical qubits, the smaller the code distance in general. 
  Its asymptotic properties are manifested that when the number of physical qubits is given $\mathcal{O}(n^3)$, the maximum number of logical qubits can be $\mathcal{O}(n)$, and the code distance is also $\mathcal{O}(n)$; if sacrificing the number of  logical qubits to be a constant, the code distance can reach $\mathcal{O}(n^2)$.
\section{Decoding Chamon model}\label{sec_decode}
  Our goal is to use the Chamon model for fault-tolerant quantum computing.
  In this section, let us introduce noise to Chamon models and propose an error correction algorithm.
  
  The errors influence of the Chamon model is complex.
  If analyzing the errors of the measurement qubits and the errors propagation between the measurement qubits and the data qubits, we even cannot recover from the errors.
  Thus, here only analyze the memory errors in Chamon model with the perfect measurement process.
  
  \subsection{Noise model}
  For analysing the memory errors on each data qubit,  naturally introduce depolarized channel to Chamon models.
Suppose the error probability of data qubits is $p$, 
  then the representation for depolarized channel is
  \begin{align}
      \notag\mathcal \epsilon(\rho)=&\frac{p^\prime I}{2}+(1-p^\prime)\rho\\
      =&(1-p)\rho+\frac{p}{3}(X\rho X+Y\rho Y+Z\rho Z),
    \end{align}
  where $p=3p^\prime/4$.
  This channel includes $X$ flips, $Y$ flips and $Z$ flips and all the  error rates are equal to $p/3$.
  
  To obtain the error correction capability in Chamon models, it is necessary to consider the meaning of the logical error rate and threshold error rate. 
  The success or failure of an error correction process is accidental, which is not suitable for measuring the error correction ability of stabilizer codes. 
  Usually, use the Monte Carlo method to estimate the  error probability of logical qubits. 
  If $N$ experiments are performed and the number of decoding failed is $N_r$, then the logical error rate is defined as
  \begin{align*}
    P_L=f(p)=\lim_{N\to \infty} N_r/N.
  \end{align*}
  It can be seen that $f$ is an increasing function.
  For a fixed stabilizer code, the asymptotic properties of the logical error rate $P_L$ as a function of $p$ depend on the code distance $d$.
  It shows that when $d$ is relatively large and $p$ gets smaller, error correction process has a higher probability of success, that is $P_L$ becomes smaller.
  Strictly speaking, for sufficiently large $d$, sufficiently small $p$, $P_L$ can be arbitrarily small.
  \begin{definition}
      $\forall \varepsilon>0$, $\exists$ $d_0, p^\prime_{th}$ such that
      when $d>d_0$ and $p<p^\prime_{th}$, $|P_L-0|<\varepsilon$.
      The threshold error rate is defined as $p_{th}=\sup\{p^\prime_{th}\}$.
  \end{definition}
  \subsection{Decoding algorithm}
  For estimating the threshold, a decoding algorithm is proposed to calculate logical error rate.
  The core idea of our error correction algorithm is first to eliminate errors globally  and then search the logical state space.
  Considering the local properties of data qubit
  errors, the algorithm is improved by adding a pretreatment process.
  
  The step of eliminating errors is challenging in the Chamon model, which is distinct from general topological codes.
  For example, the most common minimum-weight perfect-matching algorithm is applied widely to decode surface codes, in which any given match corresponds to one way to eliminate errors. However, the error distribution of the stabilizers in Chamon model is chaotic so that error elimination cannot be obtained directly.
  
  \subsubsection{Error elimination algorithm}
  Define a recover Pauli operator, denoted by $P_{\rm r}$, which can eliminate all measurement errors.
  In this section, a kind of expression of $P_{\rm r}$ is obtained by the proposed error elimination algorithm.
  Though it may not help to explore the right logical state, the errors of the stabilizers are eliminated.
  
  The algorithm can be listed as $4$ steps, which suits the case when $g(\alpha_x,\alpha_y)=1$.
  
  \textbf{Step 1.} 
  The first step is to apply $Z$-flips in the $2\alpha_z$ planes perpendicular to the $z$ axis, and the order is towards the positive $x$ axis.
  The details are as follows. 
  In each plane, denoted briefly by $z=z_i$, $x=x_j$ starts from $0$ to $2\alpha_x-3$, each time $x_j$ increasing one each time, $i=0, \cdots, 2\alpha-1$.
  Every fixed $z=z_i$ and $x=x_j$ forms a line with $\alpha_y$ stabilizers.
  The coordinates of these stabilizers are related to $x, z$. 
  Start from the stabilizer with the minimum $y$ coordinate and observe the measurement result.
   If the  measurement result of the $k$-th stabilizer, the coordinate denoted by $s_{i,j,k}$, changes on this line, apply $Z$ flip to the nearest data qubit in the positive $x$-axis direction, denoted by $Z_{s_{i,j,k}+e_x^{+}}$. 
   The measurement results need to be updated after each operation. Last, record all qubits $Z$-flips and get
  \begin{align*}
       P_1=\otimes Z_{s_{i,j,k}+e_x^{+}}
  \end{align*}
  This step makes the errors restricted to the plane of $x=2\alpha_x-2$ and $x=2\alpha_x-1$.
  
  \textbf{Step 2.} 
  Apply  $X$-flips in the remaining two planes in Step 1, and the order is towards the positive $z$ axis.
  Specifically, for any $x=x_j$, $z=z_i$ starts from $0$ to $2\alpha_z-3$,$j=2\alpha_x-2, 2\alpha-1$.
  Similar to Step 1, in this way $z=z_i$ and $x=x_j$ forms a line with $\alpha_y$ stabilizers.
  Start from the stabilizer with the minimum $y$ coordinate and observe the measurement result.
  If the measurement of the $k$-th stabilizer changes, apply $X$ flip to the nearest data qubit in the positive $z$-axis direction and update the measurement results.
  Record the whole bit flips, that is
  \begin{align*}
       P_2=\otimes X_{s_{i,j,k}+e_z^{+}}
  \end{align*}
  This step makes the errors distributed in the stabilizers set $\mathcal S|_{\{x=2\alpha_x-2,2\alpha_x-1\}\cap\{z=2\alpha_z-2,2\alpha_z- 1\}}$.
  Intuitively, the errors are restricted to four lines parallel to the $y$ axis.
  
  \textbf{Step 3.} 
  Eliminate errors on every line.
  Before the process of Step 3, we supplement some properties.
  Similar to data qubits, the stabilizers can also be classified into $4$ classes, denoted by $\mathcal S^{s_i}$,
  \begin{small}
    \begin{align}
      \mathcal S^{s_i}=\{S_s|s=s_i+2\lambda_x e_x^{+}+2\lambda_y e_y^{+}+2\lambda_z e_z^{+},\notag\\(\lambda_x,\lambda_y, \lambda_z)\in\mathcal Z_{\alpha_x}\times\mathcal Z_{\alpha_y}\times\mathcal Z_{\alpha_z}\},
    \end{align}
\end{small}
    where $
      s_0=(0,1,0),s_1=(1,0,0),
      s_2=(0,0,1),s_3=(1,1,1).
      $
    In  Chamon models, one can check that 
    \begin{lemma}\label{lem_22222}
     For any $i=0,1,2,3$, the number of the stabilizers with changed measurements in $S^{s_i}$ is even.
  \end{lemma}
  Note that all the stabilizers on the line parallel to the $y$ axis belong to exactly one class of stabilizers.
  By Lemma \ref{lem_22222}, the number of stabilizers whose measurements change on each line is even.
  
  Before explaining the process of Step 3, there is one property left to prove.
  On the line parallel to the $y$-axis, consider the two stabilizers $S_{s_{(1)}},S_{s_{(2)}}\in \mathcal S^{s_i}$. Define two stabilizers  
  the second-nearest neighbor stabilizers, if they obey
  \begin{align*}
    s_{(1)}-s_{(2)}=4e_y^{\pm}.
  \end{align*}
  Then another useful property is shown as follows:
  \begin{lemma}\label{lem_cijinlin}
    Any two second-nearest neighbor stabilizers are denoted as $S_{s_{(1)}}$ and $S_{s_{(2)}}$, and the midpoint of $s_{(1)}$ and $s_{(2)}$ is denoted as $s_{(0)}$.
    In the Chamon model, if $\alpha_x, \alpha_y$ are coprime, then there exists some $Z$ flips, whose support parallel to the $xOy$ plane, denoted as $P_Z(s_{(0)})$, so that the measurements of $S_{s_{(1)}}$ and $S_{s_{(2)}}$ are changed, and all else are unchanged.
  \end{lemma}
  
  The proof of Lemma \ref{lem_cijinlin} is put in \ref{sec_appendix_lemma}.
  Here let us set an example illustrating the idea of this proof.
  \begin{figure}
    \centering
    \subcaptionbox{\label{fig_chainjifa-12}}
    {
    \includegraphics[width=0.3\linewidth]{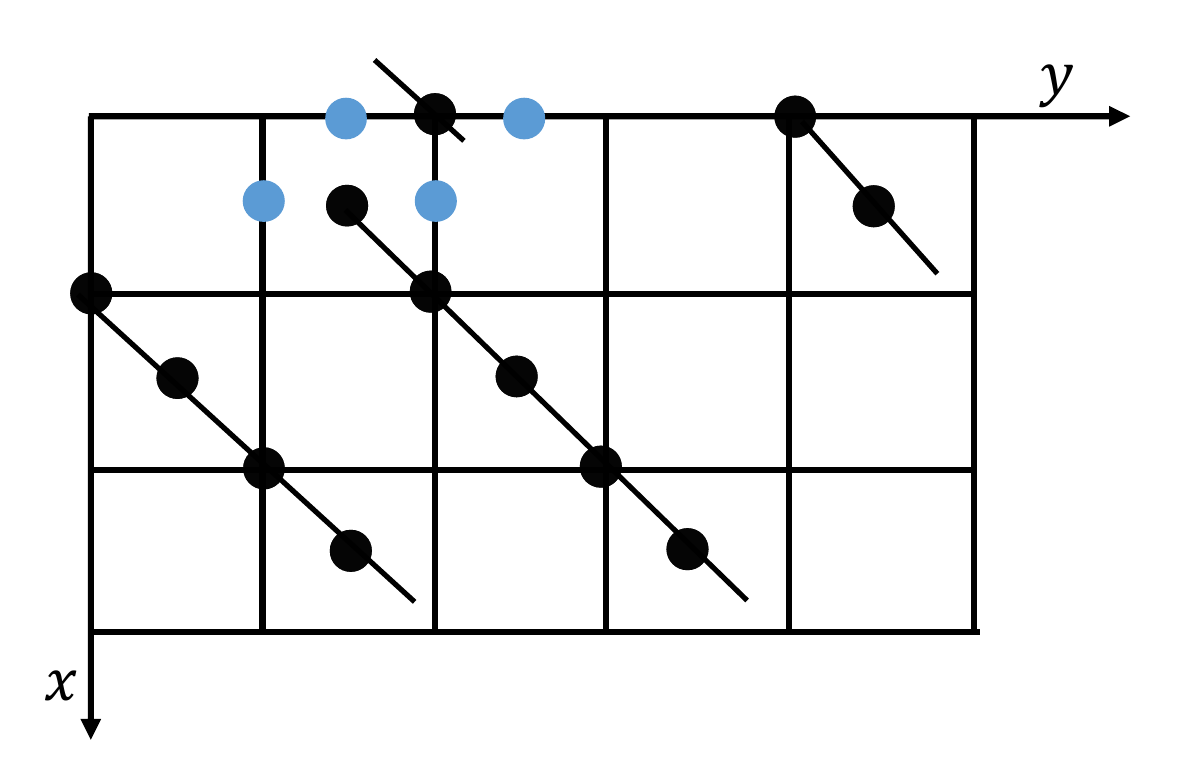}
    }
    \subcaptionbox{\label{fig_chainjifa-13}}
    {
    \includegraphics[width=0.3\linewidth]{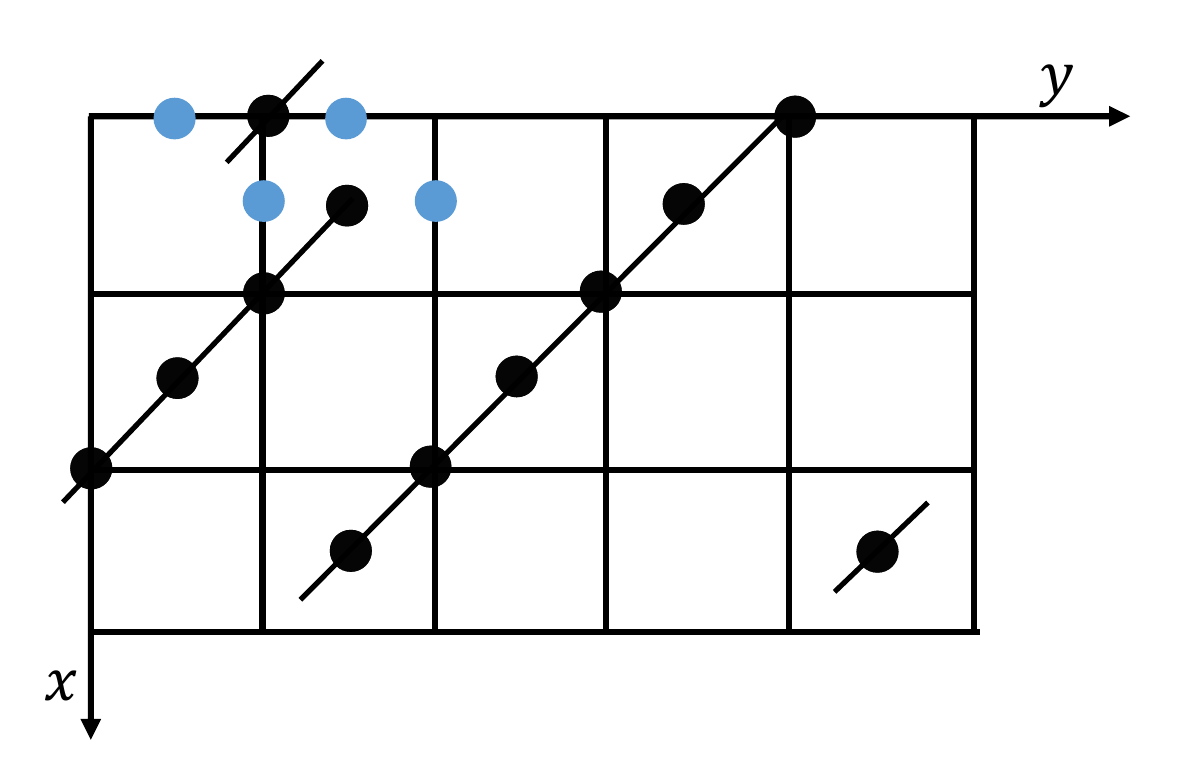}
    }
    \subcaptionbox{\label{fig_chainjifa-14}}
    {
    \includegraphics[width=0.3\linewidth]{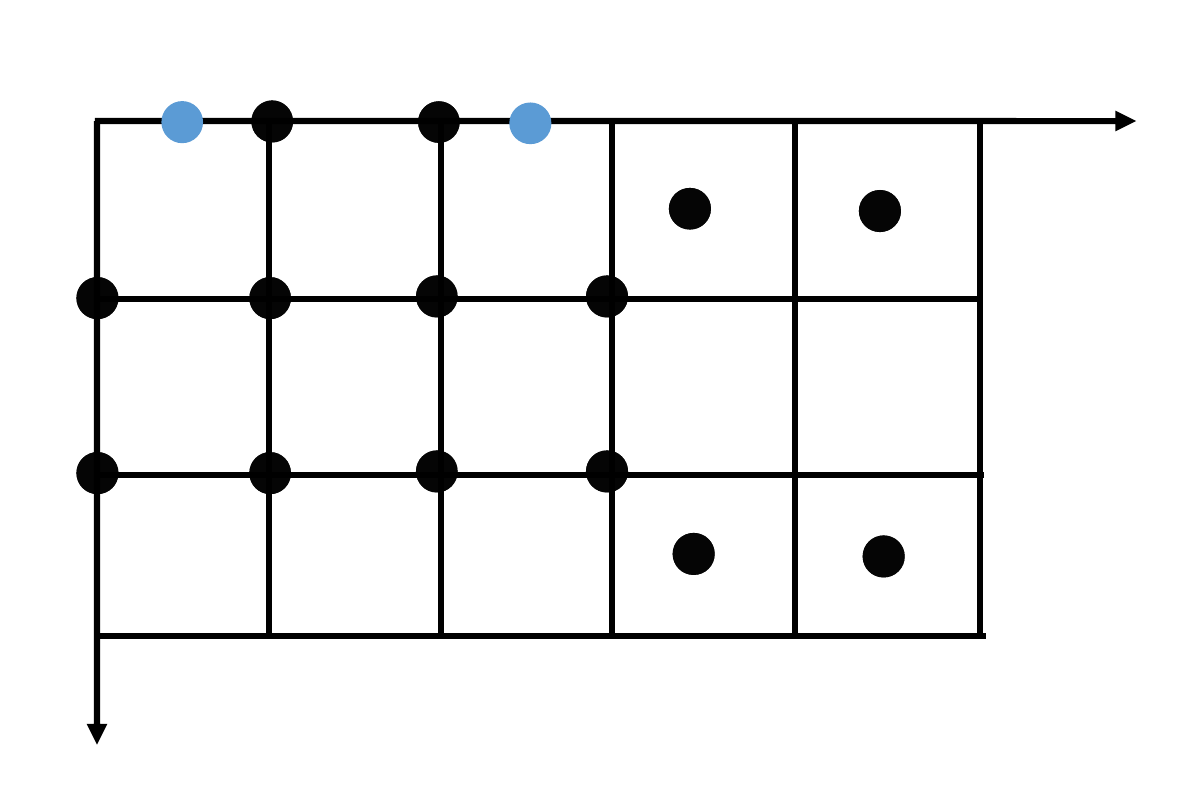}
    }
    \caption{{\bf The product of chain $Z$ operators changes the outcomes of the second-nearest neighbor stabilizers}.
    (a-b) In the direction of $e_x^{+}+e_y^{-}$ or $e_x^{+}+e_y^{+}$, the chain $Z$ operator, whose support colored black, changes only $4$ adjacent stabilizers, colored blue.
    (c)The product of two chain $Z$ operators, denoted by $P_Z$, guarantees the outcomes change of the second-nearest neighbor stabilizers.
     \label{fig_chainjifa}}
    \end{figure}
\begin{example}\label{exa_chainexcit}
        Set $\alpha_x=3$, $\alpha_y=5$, as shown in Fig.$\ref{fig_chainjifa}$.
        Consider four adjacent stabilizers, belonging to two classes of stabilizers.
        There is a chain $Z$ operator, acting on the data qubits arranged diagonally in the direction of $e_x^{+}+e_y^{+}$, which can just change the outcomes of these stabilizers, as shown in Fig.$\ref{fig_chainjifa-12}$. The support has $12$ data qubits.
        Similarly, consider the symmetrical case, as shown in Fig.$\ref{fig_chainjifa-13}$.
        The chain $Z$ operators in the direction of $e_x^{+}+e_y^{-}$ can also change the measurement results of the four adjacent stabilizers.
        Finally, the product of these two chain operators results in the changes of the second-nearest neighbor stabilizers measurements, as shown in Fig.$\ref{fig_chainjifa-14}$.
      \end{example}
      Now start to describe Step 3. 
      On each line formed by $\alpha_y$ stabilizers, sort the stabilizers along the $y$ axis, from $0$ to $\alpha_y-1$.
      Consider the $k$-th stabilizer, whose coordinate denoted by $s_k$, $k=0,\alpha_y-3$. $k$ increases sequentially from $0$, and if the $k$-th stabilizer measurement changes, then apply $P_Z(s_k+2e_y^+)$. 
      This step recovers the measurements of the stabilizers located at the $0$-th to the $(\alpha_y-3)$-th.
      That is, on each line there are at most two points with errors, whose coordinates are  $s_{\alpha_y-2}$ and $s_{\alpha_y-1}$. 
      
      Before Step 4, briefly analyze the situation of the two stabilizers measurements on each line.
      Two stabilizers belong to one class and by Lemma $\ref{lem_22222}$ the measurement results either both change or both have not changed.
      If the stabilizer measurements do not change, the errors on this line have been eliminated, or perform Step 4.
      
      \textbf{Step 4.} 
      Repeat Step 3 from $s_{\alpha_y-2}$ with step-length 2.
      
      The reasons for eliminating all errors are as follows.
      Observing the $(\alpha_y-2)$-th stabilizer, if the measurement result changes, the measurement result of the $(\alpha_y-1)$-th stabilizer also changes. Therefore, according to Step 3, apply the second-nearest neighbor operator $P_Z(s_{\alpha_y-1})$.  The $0$-th and the $(\alpha_y-1)$-th two stabilizers measurement results still change. Since $\alpha_y-1$ is an even number, thus all errors are eliminated by applying a series of the second-nearest neighbor operators $ P_Z(s_{\alpha_y-2}) \cdots P_Z(s_{3})\cdot P_Z(s_{1}) $.
      
\begin{figure}
          \centering
          \subcaptionbox{\label{fig_chain_jiucuoa}}
          {
          \includegraphics[width=0.3\linewidth]{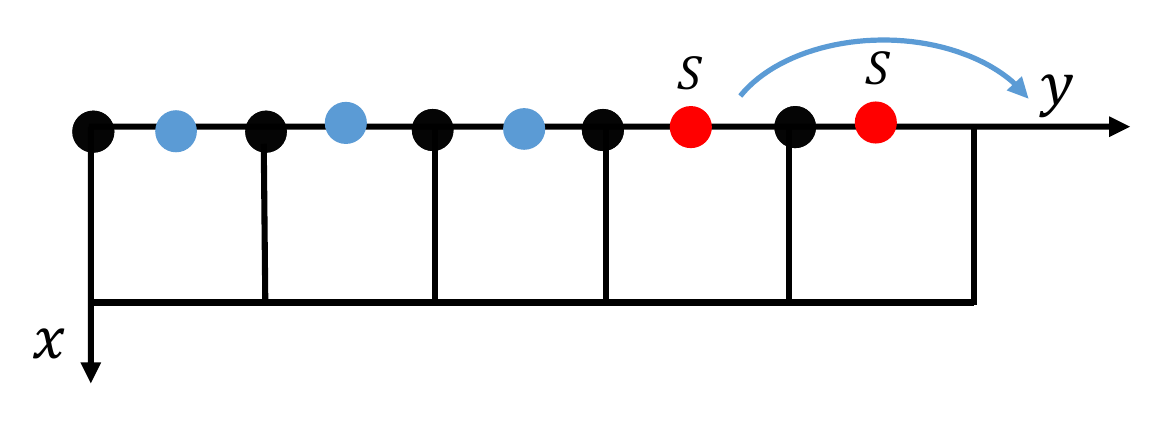}
          }
          \subcaptionbox{\label{fig_chain_jiucuob}}
          {
          \includegraphics[width=0.3\linewidth]{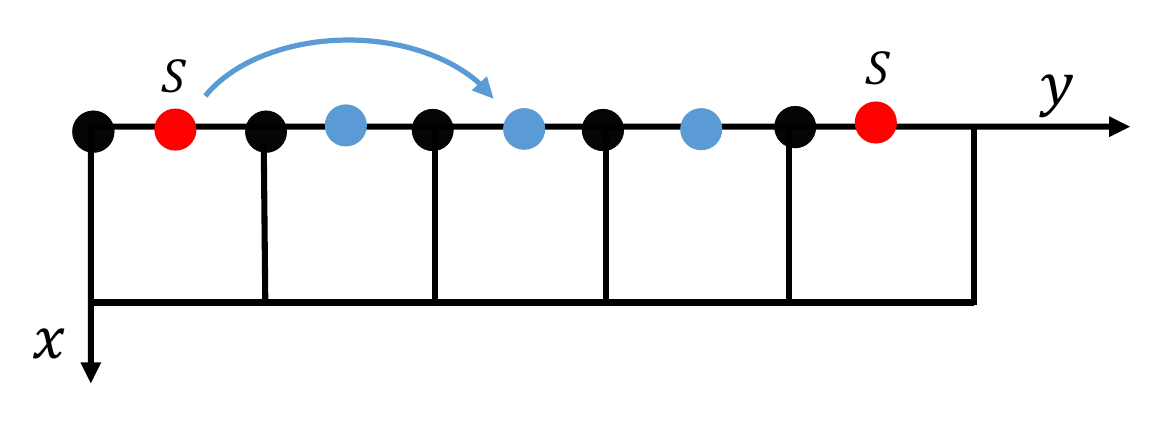}
          }
          \subcaptionbox{\label{fig_chain_jiucuoc}}
          {
              \includegraphics[width=0.3\linewidth]{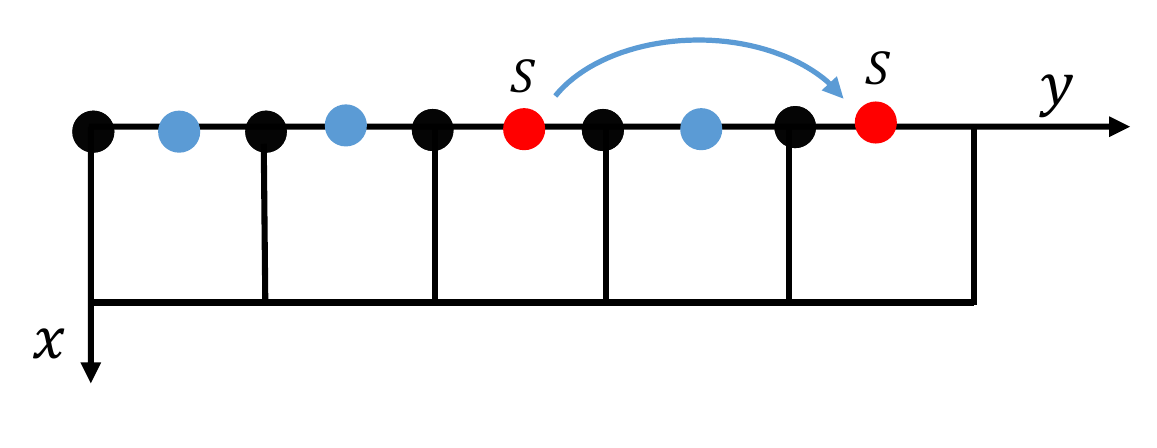}
          }
          \caption{{\bf Step 4 of error elimination algorithm}.
          (a) Translate the error in  $s_3$ to $s_0$ by applying $P_Z(s_4)$;
          (b-c) Translate the error in  $s_0$ to $s_4$ by applying $P_Z(s_3)\cdot P_Z(s_1)$.
           \label{fig_chain_jiucuo}}
          \end{figure}

      Here is an example to illustrate the process in Step 4.
      \begin{example}
          Considering a Chamon model with $\alpha_y=5$, after the first three steps, suppose there are two errors on a line parallel to the $y$ axis.
          These two errors are on the $k$-th stabilizer, $k=3,4$, as shown in Fig. $\ref{fig_chain_jiucuoa} $.
          After the second-nearest neighbor Pauli $Z$ operator  $ P_Z(s_{4})$, the errors are distributed in the $0$-th and the $4$-th stabilizers, as shown in Fig. $\ref{fig_chain_jiucuob} $.
          The error propagates from the $0$-th stabilizer to the $k$-th stabilizer, where $k$ is even, until $k=\alpha_y-1=4$. Thus, errors on this line are eliminated, as shown in Fig. $\ref{fig_chain_jiucuob} $ and Fig. $\ref{fig_chain_jiucuoc}$.
      \end{example}
      
          Denote the product of all the second-nearest-neighbor Pauli $Z$ operators in Step 3 and Step 4 as $P_{34}=\prod_{i}P_Z(s_{i})$.
          The error elimination algorithm shows that the recover Pauli operator is $P_{\rm r}=P_{34}P_2P_1$.

      Note that this algorithm requires that there exist two dimensions coprime, so the algorithm has a wide range of applications to Chamon models.
      In the construction process of Chamon model, given the scales of the total data qubits, such as $\mathcal O(n^3)$, one can build a Chamon model in which  three dimensions are $\mathcal O(n)$ and pairwise coprime.
      
      \subsubsection{Global randomized  error correction algorithm}
      In this section, the global randomized error correction algorithm is proposed for Chamon models with $\alpha_x,\alpha_y,\alpha_z$ pairwise coprime.
      In a numerical experiment, the basic idea is first to randomize the error elimination algorithm $L$ times to obtain corresponding recover Pauli operators, denoted by $P_{{\rm r}l}$, $l=0,\cdots,L-1$. 
      Then apply half-plane logical operators $X|_{\mathcal D_x^{g_i}}$ and $Z|_{\mathcal D_z^{g_i}}$ to $P_{{\rm r}l}$ and choose the final recover Pauli operator $P_{\rm r}$ with the minimum weight.
      
      The process of randomization is described in three ways.\\
        (1) The starting point is chosen randomly. 
        In Step 1, randomize the starting data qubits. 
        The original algorithm starts from the data qubit located at $(0,0,0)$. In fact, any position of data qubit can be used as a starting point. After $P_1$, the errors are included on two adjacent planes. 
        In Step 2, similarly, randomize the starting data qubits. 
        The same idea suits Step $3-4$. 
        This is all caused by the translation invariance of the Chamon model as a 3D torus.\\
        (2) The order of Step 1 and Step 2 can also be adjusted.
        One can apply $X$ operators first, and then follow the $Z$ operators.\\
        (3) From Step 3 to Step 4, performing the second-nearest neighbor Pauli $Z$ operator needs to limit $\alpha_x$ and $\alpha_y$ coprime. Similarly,
      Since $\alpha_y,\alpha_z$ are coprime, the second-nearest neighbor $X$ operator can be selected.
      
\begin{figure}
        \centering
        \includegraphics[width=0.8\linewidth]{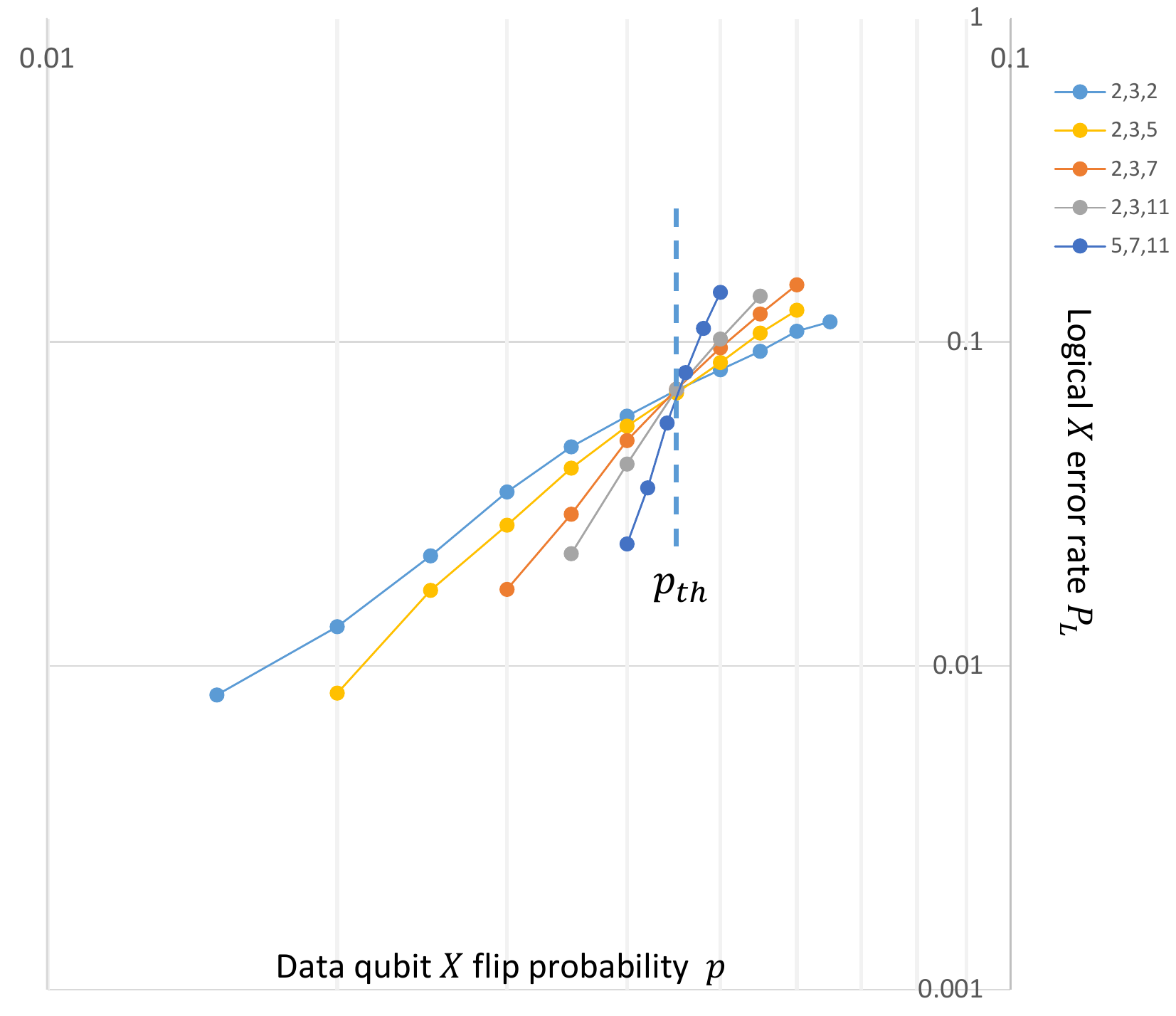}
        \caption{{\bf Logical $X$ error rate $P_L$ as a function of data qubit $X$-flip
        probability $p$. } Here shows various Chamon models denoted by $\alpha_x,\alpha_y,\alpha_z$ with different colors and the threshold error rate is $4.45\%$ for the global error correction algorithm.
         \label{fig_whole_pth}}
        \end{figure}

      Finally, let us summarize the global randomized algorithm.
      Firstly, introduce the standard depolarizing channel to Chamon models with perfect measurements. 
      Then apply the recover Pauli operator $P_{\rm r}$ to correct the data qubits errors.
      For each recovered state, test for logical $X$ failure by checking whether there are an odd
      or even number of $X$ flips on any planes perpendicular to the $z$-axis.
      There will be an odd number of $X$-flips in the plane perpendicular to the $z$-axis if a half-plane logical $X$ operator has been applied, indicating a logical $X$ error.
      By performing many simulations, the probability of logical $X$ error $P_L$ versus $p$ can be plotted, as shown in Fig. \ref{fig_whole_pth}.
      The threshold error rate is about $4.45\%$.
      The dimensions of the three models denoted by $(\alpha_x,\alpha_y,\alpha_z)$ are $(2,3,2)$, $(2,3,5)$, $(2,3,7)$, $(2,3,11)$ and $(5,7,11)$, respectively.
      
      Note that the algorithm does not reflect the impact of local qubits errors, but considers all stabilizers measurements. Thus, this error correction algorithm is global.
      A natural question is whether the error correction capability of the Chamon model can be improved when  local errors form is considered.
      Next, introduce this idea to the Chamon models and find the threshold can be improved slightly.
      
      \subsection{Improved error correction algorithm}
      In this section, the global error correction algorithm is improved by a pretreatment process, termed the probabilistic greedy local algorithm.

      Although the global algorithm can eliminate errors definitely, when recovering the original logical state, it may be led to randomness due to global searching.
      The basic idea for the probabilistic greedy local algorithm is to decrease the number of false stabilizers greedily by using the property that any qubit memory error can affect the nearest $4$ stabilizers. 
      Details of specific procedures are put in \ref{sec_appendix_greedy}. 
      Hence the process of the improved algorithm is listed as follows:
      \begin{itemize}
      \item[\textbf{(a)}] Execute the probabilistic greedy local algorithm. If the errors are all eliminated, skip to Step (c).      
      \item[\textbf{(b)}] Execute the global randomized error elimination algorithm.
      \item[\textbf{(c)}] Apply half-plane logical operators $X|_{\mathcal D_x^{g_i}}$ and $Z|_{\mathcal D_z^{g_i}}$ to search the recover Pauli operator with the minimum weight.
      \item[\textbf{(d)}] Use parity check method to compute $P_L$.
      \end{itemize}
      
\begin{figure}
          \centering
          \includegraphics[width=0.8\linewidth]{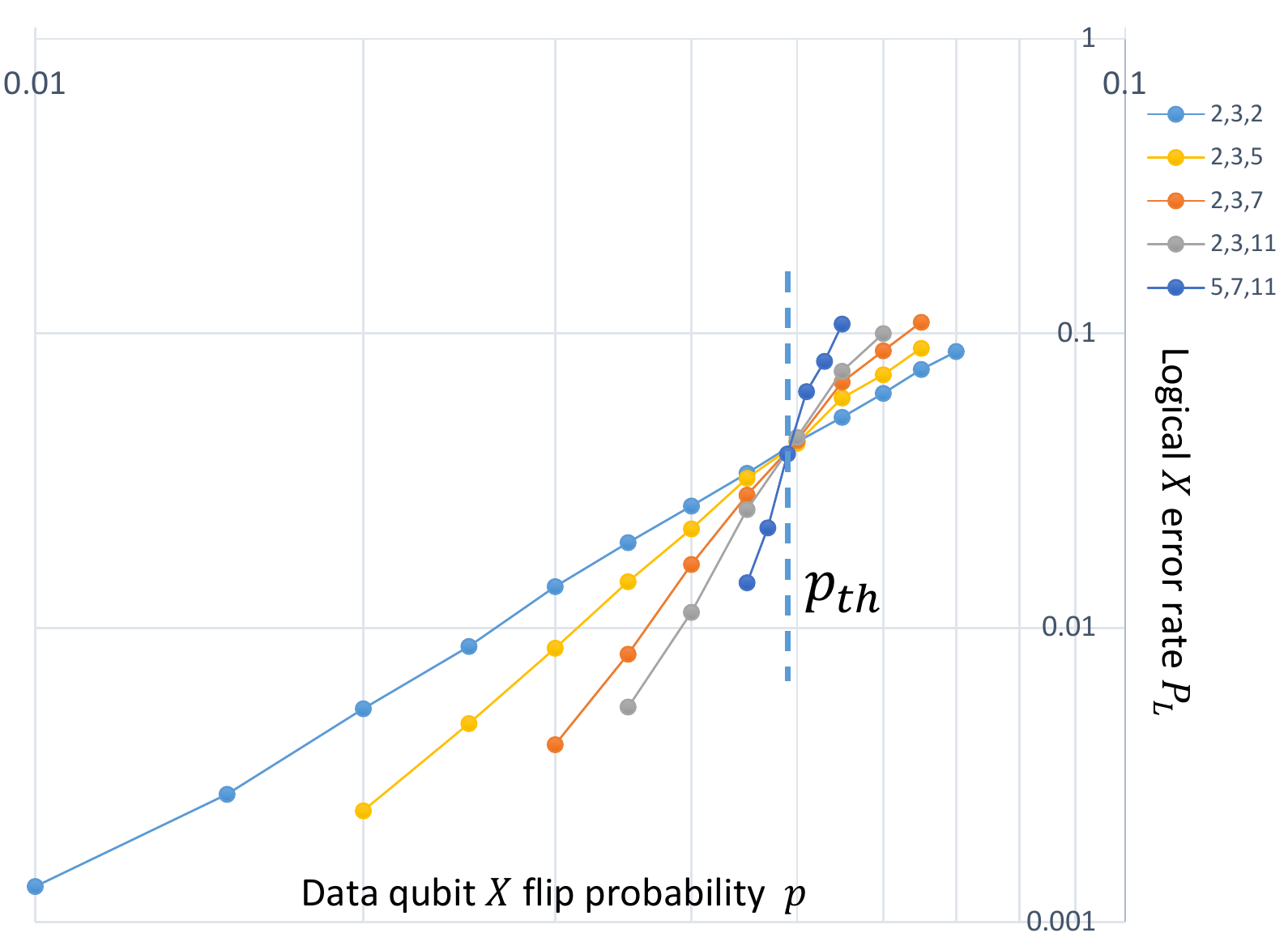}
          \caption{{\bf Logical $X$ error rate $P_L$ as a function of $X$-flip
          probability $p$.} Chamon models in different dimensions are denoted by $\alpha_x,\alpha_y,\alpha_z$ with different colors. The threshold error rate for the improved error correction algorithm is about $4.92\%$.
           \label{fig_mix_pth}}
          \end{figure}

      Using the improved error correction algorithm, the probability of logical $X$ error $P_L$ versus $p$ can be plotted, as shown in Fig. \ref{fig_mix_pth}.
      The threshold error rate of this algorithm is about $4.92\%$, slightly higher than that computed by the global error correction algorithm.
      
        It is clear that the improvement of the local error correction algorithm to the global algorithm is not essential.
          When $p$ is small, for a given data qubit error rate $p$, it can be found that the logical error rate $P_L$ of the improved error correction algorithm will be less.
          However, around the threshold, the two algorithms exhibit essentially similar performance.
          The reason is that the effect of the stabilizers is excluded from consideration in the process of recovering Pauli operators.
          However, in the consideration of stabilizers, decoding becomes an exponentially searching problem. The improvement of the decoding algorithms for Chamon models is also our future research direction.
\section{Conclusion}\label{sec_conclusion}
The Chamon model in this paper becomes a kind of error correction code.
Compared to previous work, this paper thoroughly explains how $4\alpha$ logical qubits are constructed. The logical operators and the related properties are given for any Chamon model. The threshold for Chamon model is obtained under the perfect measurements.

In the decoding process, the proposed probabilistic greedy local algorithm adapts to different kinds of high-dimensional models.
And the advantage of this retreatment process lies in its low cost without consideration of global properties for error correction codes.
A possible future research topic is to generalize the error correction algorithm to other high-dimensional codes.
  
 The global decoding algorithm depends on the linear dimensions for Chamon models, which implies topological properties in Chamon models.
 It cannot be suitable for the case that any two dimensions are not coprime.
 Hence it remains to be studied  how to generalize the applicable situations.
 On the other hand, the local error correction algorithm also has room for improvement.
 For example, one can design an algorithm more exquisitely by considering the second-nearest interactions to improve the performance of the algorithm. 
\section*{Acknowledgements}

 This work was supported by the National Natural Science Foundation of China (Grant No. 12034018) and Innovation Program for Quantum Science and Technology
 (Grant No. 2021ZD0302300).

\appendix

\section{Chamon model in toric representations}\label{sec_appendix_torus}
In this section, the Chamon model is represented in a torus. The support of logical operators in this representation is a non-contractible loop on the torus.
The stabilizers can be regarded as a circle with some discrete points.
Our goal is to explore the shortest loop on the torus, corresponding to the logical operator with the minimum weight.

\subsection{Stabilizers and data qubits in toric representations\cite{bravyi2011topological}}

\begin{figure}[h]
  \centering
  \includegraphics[width=0.5\linewidth]{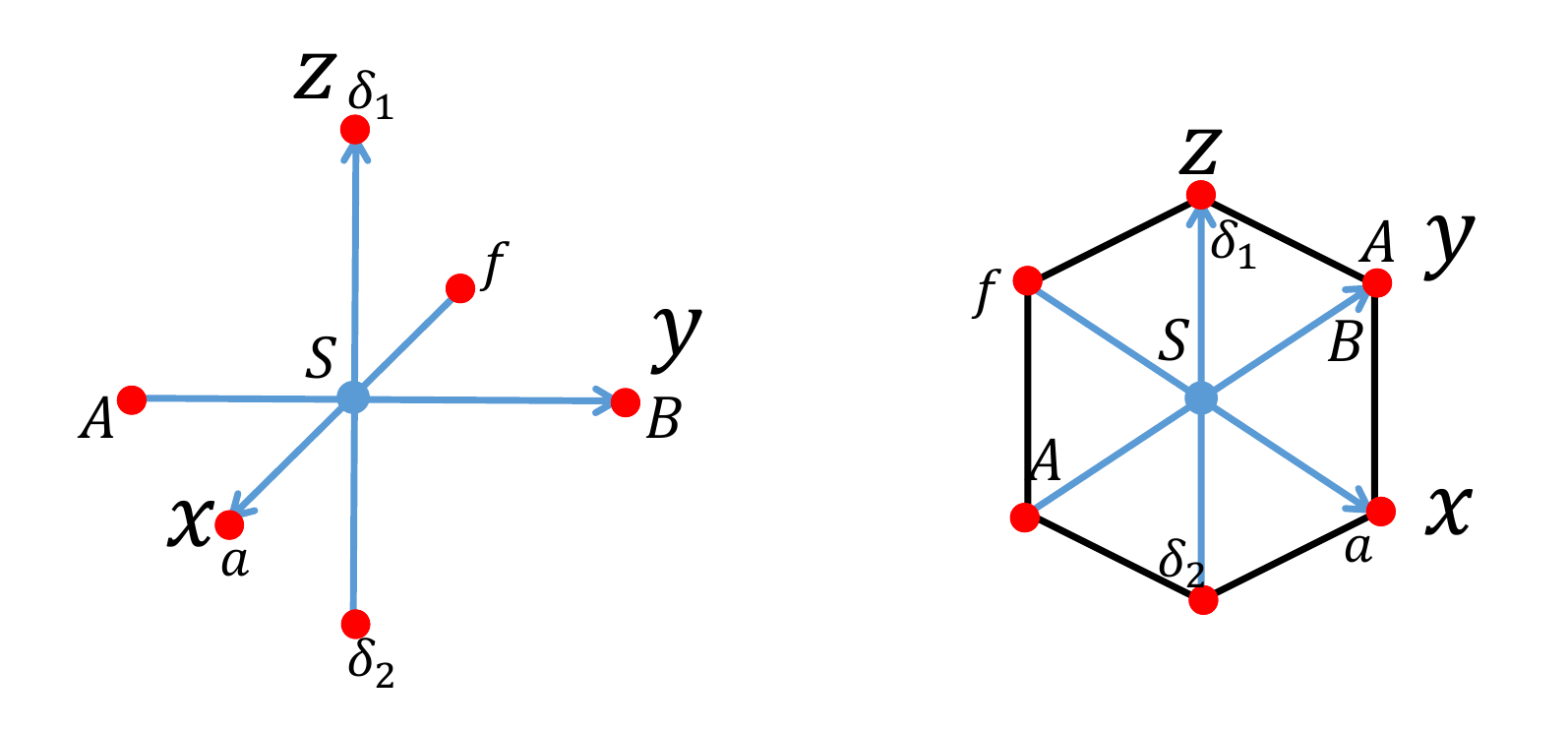}
  \caption{{\bf The stabilizer and corresponding hexagon representations}.
  $S=X_aX_fY_AY_BZ_{\delta_1}Z_{\delta_2}$.
   \label{fig_plain_definition}}
\end{figure}

First, introduce the idea of the representations of stabilizers.
The stabilizer and the corresponding 6 data qubits are represented by a regular hexagon, in which the center represents the stabilizer, and each vertex represents data qubits. A regular hexagon has three longest diagonals, whose two endpoints represent the same measurement type, forming the three positive directions of $x,y,z$, as shown in Fig. \ref{fig_plain_definition}.

\begin{figure}[h]
  \centering
  \subcaptionbox{}
  {
  \includegraphics[width=0.3\linewidth]{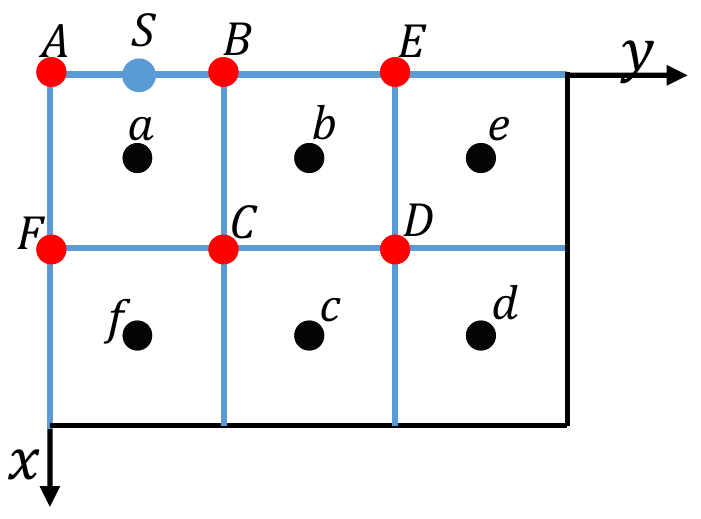}
  }\\
  \subcaptionbox{\label{fig_xoy_6b}}
  {
  \includegraphics[width=0.8\linewidth]{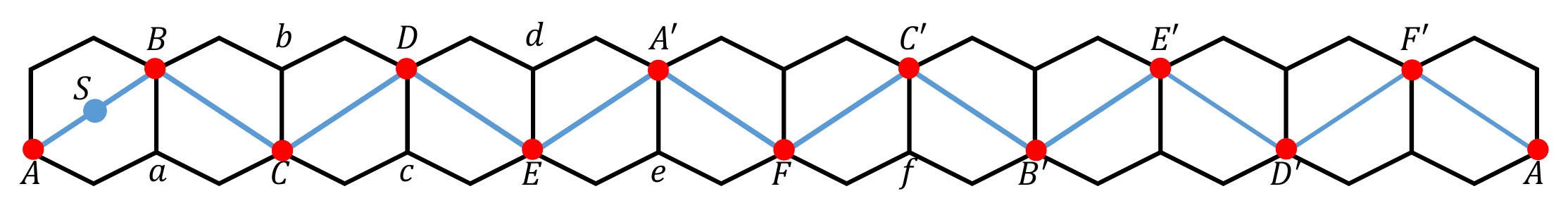}
  }
  \caption{{\bf The representations of stabilizers and data qubits on $xOy$ plane}.
  (a) Data qubits are marked with dots and letters, and stabilizers are represented by blue edges. The data qubits used by the half-plane logical $Z$ operator in the $xOy$ plane are marked with different colors. 
  (b) Stabilizers are represented by the centers of regular hexagons and data qubits are represented by vertices.  
  \label{fig_xoy_6}}
  \end{figure}

Then consider the stabilizers on the $xOy$ plane.
Note that these stabilizers on the $xOy$ plane can be arranged diagonally in the $e_x+ e_y$ direction. 
It can be obtained that every two adjacent stabilizers share two data qubits. 
Suppose $\alpha_x$ and $\alpha_y$ are coprime, then these diagonally arranged stabilizers fill the whole $xOy$  plane under the periodic boundary conditions. 
In the $e_x+e_y$ direction, two adjacent stabilizers share two data qubits, corresponding to two regular hexagons sharing an edge. 
Fig. \ref{fig_xoy_6} shows hexagon representation when $\alpha_x = 2, \alpha_y = 3$. 
Note that one data qubit corresponds to two vertices of the regular hexagons, 
because each data qubit on the plane is coupled with four stabilizers, and its vertex is shared by two regular hexagons. Thus the data qubits and stabilizers on the $xOy$ plane can be represented by a row of regular hexagons.

  Last, generalize the method on $xOy$ plane to the entire Chamon model. 
  A natural idea is that the data qubits and the stabilizers in each plane perpendicular to the $z$-axis can be represented as a row of hexagons similarly. Thus intuitively there are $2\alpha_z$ rows of regular hexagons. 

  In this tessellation way, each vertex shares three hexagon faces, and each face shares 6 vertices. 
  This is inconsistent with the data qubits being measured by the nearest 6 stabilizers, which is twice the number of shared faces.
  Therefore, a straightforward and concise implementation is just to map each data qubit to two vertices. Due to the periodic boundary conditions of directions of $e_z$ and $e_x+e_y$, mathematically the Chamon model can be represented as a torus. Here illustrate the toric representation of a Chamon model with an example.

\begin{figure}[h]
  \centering
  \includegraphics[width=1\linewidth]{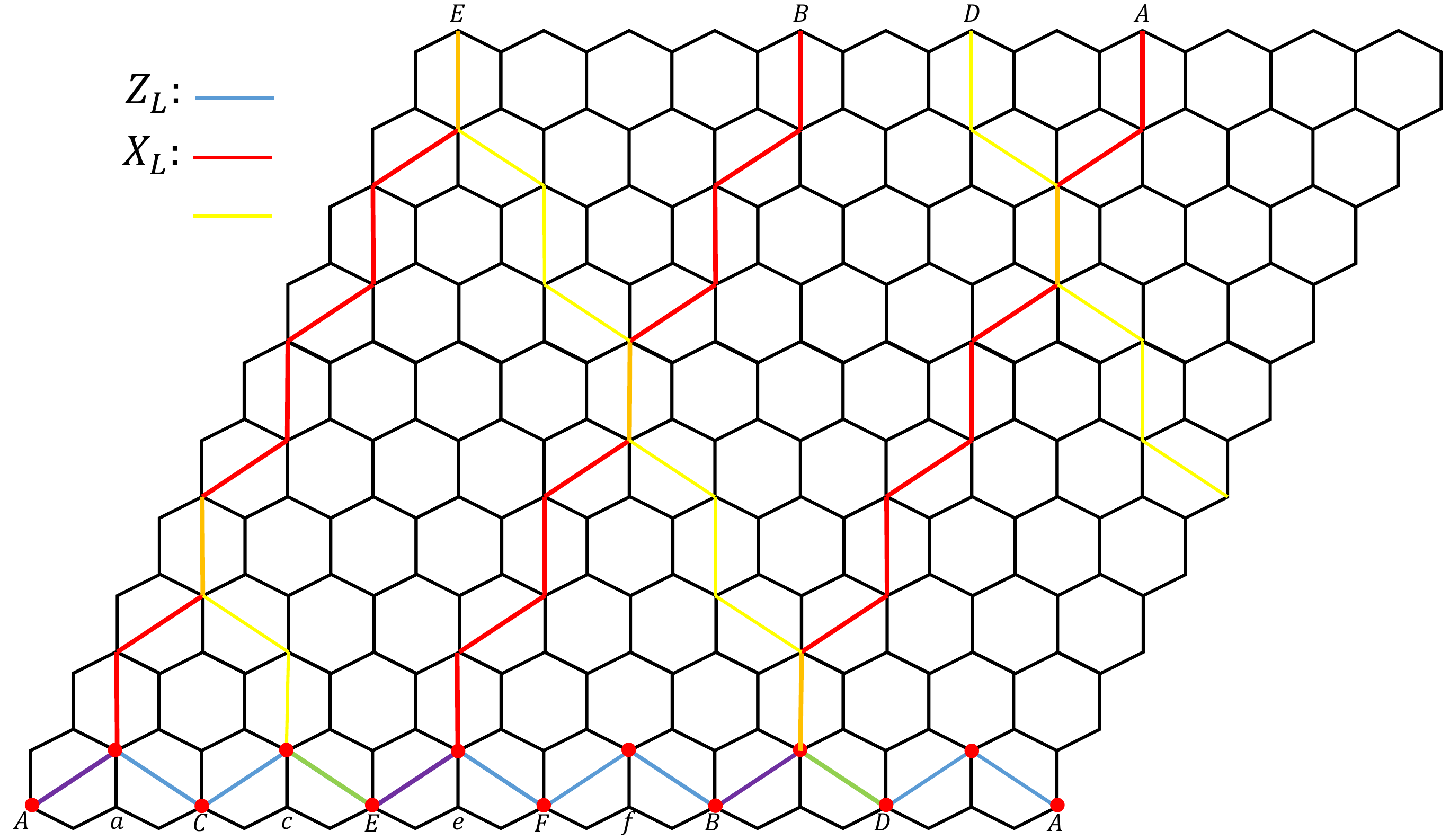}
  \caption{{\bf The Chamon model in toric representations}.
  The logical operators are displayed in three color chains, in which the blue chain and red chain  represent the logical $Z$ operator and logical $X$ operator, respectively. The yellow one depends on the parity of $\alpha_x$ and $\alpha_z$.
  The composite of colors indicates the overlapping of chains.
   \label{fig_plain_whole}}
  \end{figure}

\begin{example}
  Given a Chamon model with $\alpha_x=2,\alpha_y=3$ and $\alpha_z=5$, as shown in Fig.\ref{fig_plain_whole},
  there are a total of $2\alpha_x\alpha_y$ generators of stabilizers in the plane $z=i$, corresponding to a row of $2\alpha_x\alpha_y$ regular hexagons, $i=0,\cdots,2\alpha_z- 1$.
   When $i=0$, the data qubits on the plane inherit the notation of Fig.$\ref{fig_xoy_6}$.
   There are a total of $2\alpha_z=10$ layers, thus a total of $4\alpha_x\alpha_y\alpha_z$ regular hexagons.

\end{example}

\subsection{Logical Pauli operators in toric representations}
Suppose $\alpha_x,\alpha_y$ and $\alpha_z$ are pairwise coprime for simplicity.
Recall the half-plane logical $Z$ operator $Z_{Li}=Z|_{\mathcal D_z^{g_i}}$, the definition of $\mathcal D_z^{g_i}$ as in Eq.\eqref{eq_D_half}.
As in Fig.\ref{fig_xoy_6}(b), in toric representations, the half-plane logical operator forms a loop non-contractible.
A trivial chain is the $Z$ operator acting the data qubits arranged diagonally, then obtain $Z_L=Z_AZ_aZ_CZ_c\cdots$, that is $Z_{L0}Z_{L1}$.
This kind of full plane logical operator, termed the chain operator in toric representations, also forms a non-contractible loop.

Now give the proof of the Thm.\ref{thm_d_coprime}.
Our goal is changed to explore the loop with the minimum weights.
Similarly, the half-plane logical $Z$ operator is called a half-chain operator,
coupled with half of data qubits. In the toric representations, the length of this chain is $2\alpha_x\alpha_y$. 
In this direction, any chains with less than this length will break and fail to form a loop.
Due to each data qubit corresponding to two points on the chain, the weight of this logical Z operator is $\alpha_x\alpha_y$. 
As in Fig.\ref{fig_plain_whole}, there are two other direction half-chain operators.
It can be proved that one data qubit corresponds to two points in each chain, and the weights of the logcial operators are $\alpha_y\alpha_z$ and $\alpha_z\alpha_x$, respectively when $\alpha_x,\alpha_y$ and $\alpha_z$ are coprime.
In three directions half-chain operators are the logical operators with minimum weight in each direction. Thus, the code distance is the minimum number of $\alpha_x\alpha_y,\alpha_y\alpha_z,\alpha_z\alpha_x$ for Chamon model with pairwise coprime linear dimensions.

\section{The proof of Lemma \ref{lem_cijinlin}}\label{sec_appendix_lemma}
From translation symmetry, let $s_{(1)}=(0,1,0)$, and set $s_{(2)}=(0,5,0)$. Our goal is to find a set of data qubits on the $xOy$ plane, such that after applying $Z$ operators to all data qubits in the set, only the outcomes of $S_{s_{(1)}}$ and $S_{s_{(2)}}$ change.
    By $\alpha_x$ and $\alpha_y$ coprime, let
    \begin{align*}
        n_{xy}=\min_{n\in Z^{+}}\{n|n\alpha_x({\rm mod}\alpha_y)=1\}.
    \end{align*}
    In  Example $\ref{exa_chainexcit}$, $n_{xy}=2$.
    The geometric meaning of $n_{xy}$ in the chain $Z$ operator is the number of basic units, whose length is $2\alpha_x$.
    Then the weights of two chain Pauli $Z$ operators are $2\alpha_x n_{xy}$. Define the supports of two chain Pauli $Z$ operators as
     \begin{align*}
        \mathcal D_z^{++}=&\{d|d=s_{(0)}+e_x^{+}+\lambda (e_x^{+}+e_y^{+}),\lambda\in\mathcal Z_{2\alpha_x n_{xy}-1}\},\\
        \mathcal D_z^{+-}=&\{d|d=s_{(0)}+e_x^{+}+\lambda (e_x^{+}-e_y^{+}),\lambda\in\mathcal Z_{2\alpha_x n_{xy}-1}\}.\\
    \end{align*}
    Then the Pauli $Z$ operator satisfying the Lemma \ref{lem_cijinlin} is constructed, namely
    \begin{align*}
        P_Z(s_{(0)})=Z|_{\mathcal D_z^{++}}\cdot Z|_{\mathcal D_z^{+-}}.
    \end{align*}

    Due to the translation invariance, any two second-nearest neighbor stabilizers correspond to a Pauli $Z$ operator.
    In the proof of this lemma, not only is the existence of $P_Z(s_{(0)})$ proved but the explicit expression is obtained.

\section{The probabilistic greedy local algorithm}\label{sec_appendix_greedy}
The procedures of the greedy local algorithm for Chamon model are as follows:\\
    \textbf{Step 1. Assignment.}
    A three-dimensional vector, recorded as $w=(w_x,w_y,w_z)$, is attached to each data qubit, representing three weights, where $w_x$ is defined as the number of errors of four adjacent stabilizers located in the plane perpendicular to the $x$ axis, so $1\leq w_x\leq 4$.
    Each data qubit also corresponds to an auxiliary three-dimensional vector with initialized value, denoted by $v=(0,0,0)$.
    Similarly, $w_y$ and $w_z$  have the same meanings.\\
  \textbf{Step 2. Eliminating $w_p=4$.}
    Read the weights $w_p$ of data qubits in order, $p=x,y,z$.
    If $w_p=4$, then apply the Pauli operator $P$ to the corresponding data qubit and update the weights of itself and the adjacent data qubits.
    Return to Step 2, unless all weights of data qubits $w_p<4$.
    Check whether the stabilizers errors are eliminated.
    If they have been eliminated, the algorithm succeeds, or continue to Step 3.\\
  \textbf{Step 3. Eliminating $w_p=3$.}
  Read the weights $w_p$ of data qubits in order, $p=x,y,z$.
    If $w_p=3$, then apply the Pauli operator $P$ to the corresponding data qubit, update weights and return to Step 2.
    If all weights of data qubits $w_p<3$, then check whether the stabilizers errors are eliminated.
    If errors have been eliminated, the algorithm succeeds, or continue to Step 4.\\
    \textbf{Step 4. Eliminating $w_p=2$.}
    Read the weights $w_p$ of data qubits in order, $p=x,y,z$.
    If $w_p=2$ and $v_p=0$, then apply the Pauli operator $P$ to the corresponding data qubit, update weights, record the corresponding $v_p=1$, and return to Step 2.
    If not, claim the algorithm fails.

    If the algorithm ends at Step 2 or Step 3, the algorithm succeeds, or the algorithm fails. Especially, the greedy local algorithm cannot handle the case that
    all the weights $v_p\leq 1$.

In Step 2 and Step 3, the number of error stabilizers has been decreasing. Thus, the computational complexity is $\mathcal O(N^2)$, where $N$ is the number of data qubits. In Step 4, the record of $v$ can avoid the infinite loop of the algorithm.
And in the worst case, all the data qubits have $v_p=1$, which means Step 4 is executed $\mathcal O(N)$ times. Thus the complexity for the probabilistic greedy local algorithm is $\mathcal O(N^3)$.

 \bibliographystyle{elsarticle-num} 
 \bibliography{1214.bib}

\begin{thebibliography}{10}
\expandafter\ifx\csname url\endcsname\relax
  \def\url#1{\texttt{#1}}\fi
\expandafter\ifx\csname urlprefix\endcsname\relax\def\urlprefix{URL }\fi
\expandafter\ifx\csname href\endcsname\relax
  \def\href#1#2{#2} \def\path#1{#1}\fi

\bibitem{biamonte2017quantum}
J.~Biamonte, P.~Wittek, N.~Pancotti, P.~Rebentrost, N.~Wiebe, S.~Lloyd, Quantum
  machine learning, Nature 549~(7671) (2017) 195--202.

\bibitem{shor1994algorithms}
P.~W. Shor, Algorithms for quantum computation: discrete logarithms and
  factoring, in: Proceedings 35th annual symposium on foundations of computer
  science, Ieee, 1994, pp. 124--134.

\bibitem{monz2016realization}
T.~Monz, D.~Nigg, E.~A. Martinez, M.~F. Brandl, P.~Schindler, R.~Rines, S.~X.
  Wang, I.~L. Chuang, R.~Blatt, Realization of a scalable shor algorithm,
  Science 351~(6277) (2016) 1068--1070.

\bibitem{lloyd1996universal}
S.~Lloyd, Universal quantum simulators, Science 273~(5278) (1996) 1073--1078.

\bibitem{grover1996fast}
L.~K. Grover, A fast quantum mechanical algorithm for database search, in:
  Proceedings of the twenty-eighth annual ACM symposium on Theory of computing,
  1996, pp. 212--219.

\bibitem{terhal2015quantum}
B.~M. Terhal, Quantum error correction for quantum memories, Reviews of Modern
  Physics 87~(2) (2015) 307.

\bibitem{roffe2019quantum}
J.~Roffe, Quantum error correction: an introductory guide, Contemporary Physics
  60~(3) (2019) 226--245.

\bibitem{luo2021quantum}
Y.-H. Luo, M.-C. Chen, M.~Erhard, H.-S. Zhong, D.~Wu, H.-Y. Tang, Q.~Zhao,
  X.-L. Wang, K.~Fujii, L.~Li, et~al., Quantum teleportation of physical qubits
  into logical code spaces, Proceedings of the National Academy of Sciences
  118~(36) (2021).

\bibitem{andersen2020repeated}
C.~K. Andersen, A.~Remm, S.~Lazar, S.~Krinner, N.~Lacroix, G.~J. Norris,
  M.~Gabureac, C.~Eichler, A.~Wallraff, Repeated quantum error detection in a
  surface code, Nature Physics 16~(8) (2020) 875--880.

\bibitem{ai2021exponential}
G.~Q. AI, Exponential suppression of bit or phase errors with cyclic error
  correction, Nature 595~(7867) (2021) 383.

\bibitem{erhard2021entangling}
A.~Erhard, H.~Poulsen~Nautrup, M.~Meth, L.~Postler, R.~Stricker, M.~Stadler,
  V.~Negnevitsky, M.~Ringbauer, P.~Schindler, H.~J. Briegel, et~al., Entangling
  logical qubits with lattice surgery, Nature 589~(7841) (2021) 220--224.

\bibitem{nigg2014quantum}
D.~Nigg, M.~Mueller, E.~A. Martinez, P.~Schindler, M.~Hennrich, T.~Monz, M.~A.
  Martin-Delgado, R.~Blatt, Quantum computations on a topologically encoded
  qubit, Science 345~(6194) (2014) 302--305.

\bibitem{ryan2021realization}
C.~Ryan-Anderson, J.~Bohnet, K.~Lee, D.~Gresh, A.~Hankin, J.~Gaebler,
  D.~Francois, A.~Chernoguzov, D.~Lucchetti, N.~Brown, et~al., Realization of
  real-time fault-tolerant quantum error correction, Physical Review X 11~(4)
  (2021) 041058.

\bibitem{chiaverini2004realization}
J.~Chiaverini, D.~Leibfried, T.~Schaetz, M.~D. Barrett, R.~Blakestad,
  J.~Britton, W.~M. Itano, J.~D. Jost, E.~Knill, C.~Langer, et~al., Realization
  of quantum error correction, Nature 432~(7017) (2004) 602--605.

\bibitem{schindler2011experimental}
P.~Schindler, J.~T. Barreiro, T.~Monz, V.~Nebendahl, D.~Nigg, M.~Chwalla,
  M.~Hennrich, R.~Blatt, Experimental repetitive quantum error correction,
  Science 332~(6033) (2011) 1059--1061.

\bibitem{wootton2018repetition}
J.~R. Wootton, D.~Loss, Repetition code of 15 qubits, Physical Review A 97~(5)
  (2018) 052313.

\bibitem{campagne2020quantum}
P.~Campagne-Ibarcq, A.~Eickbusch, S.~Touzard, E.~Zalys-Geller, N.~E. Frattini,
  V.~V. Sivak, P.~Reinhold, S.~Puri, S.~Shankar, R.~J. Schoelkopf, et~al.,
  Quantum error correction of a qubit encoded in grid states of an oscillator,
  Nature 584~(7821) (2020) 368--372.

\bibitem{dennis2002topological}
E.~Dennis, A.~Kitaev, A.~Landahl, J.~Preskill, Topological quantum memory,
  Journal of Mathematical Physics 43~(9) (2002) 4452--4505.

\bibitem{castelnovo2008topological}
C.~Castelnovo, C.~Chamon, Topological order in a three-dimensional toric code
  at finite temperature, Physical Review B 78~(15) (2008) 155120.

\bibitem{bravyi2011analytic}
S.~Bravyi, J.~Haah, Analytic and numerical demonstration of quantum
  self-correction in the 3d cubic code, arXiv preprint arXiv:1112.3252 (2011).

\bibitem{bonilla2021xzzx}
J.~P. Bonilla~Ataides, D.~K. Tuckett, S.~D. Bartlett, S.~T. Flammia, B.~J.
  Brown, The xzzx surface code, Nature communications 12~(1) (2021) 1--12.

\bibitem{chamon2005quantum}
C.~Chamon, Quantum glassiness in strongly correlated clean systems: An example
  of topological overprotection, Physical review letters 94~(4) (2005) 040402.

\bibitem{bravyi2011topological}
S.~Bravyi, B.~Leemhuis, B.~M. Terhal, Topological order in an exactly solvable
  3d spin model, Annals of Physics 326~(4) (2011) 839--866.

\bibitem{shirley2023emergent}
W.~Shirley, X.~Liu, A.~Dua, Emergent fermionic gauge theory and foliated
  fracton order in the chamon model, Physical Review B 107~(3) (2023) 035136.

\bibitem{nielsen2002quantum}
M.~A. Nielsen, I.~Chuang, Quantum computation and quantum information (2002).

\bibitem{knill2000theory}
E.~Knill, R.~Laflamme, L.~Viola, Theory of quantum error correction for general
  noise, Physical Review Letters 84~(11) (2000) 2525.

\end{thebibliography}





\end{document}